\documentclass[twocolumn]{aastex631}

\usepackage{subfigure}
\usepackage{amsmath}
\usepackage{threeparttable}
\usepackage{gensymb}

\shorttitle{AASTeX v6.3.1 Sample article}
\shortauthors{M.Khademi et al.}
\graphicspath{{./}{figures/}}
\begin{document}

\title{Influence of Magnetic Fields on the Gas Rotation in the Galaxy $NGC\;6946$}
\correspondingauthor{Maryam Khademi}
%\email{greg.schwarz@aas.org, gus.muench@aas.org}
\email{maryam.khademi@ipm.ir, nasiri@iasbs.ac.ir, ftaba@ipm.ir}

\author{M. Khademi}
\affiliation{School of Astronomy, Institute for Research in Fundamental Sciences (IPM), 19395-5531 Tehran, Iran}
%\affiliation{Department of Physics, Shahid Beheshti University, G.C. \\
%Daneshjou Boulevard, District 1, 19839 Tehran,  Iran}
%\affiliation{School of Astronomy, Institute for Research in Fundamental Sciences (IPM), P.O. Box 1956836613, Tehran, Iran}
%School of Astronomy, Institute for Research in Fundamental Sciences, 19395-5531 Tehran, Iran}
%\affiliation{GEPI, Observatoire de Paris, Universite PSL, CNRS, Place Jules Janssen 92195, Meudon, France}
%{P.O. Box 19395-5746, Tehran - Iran}

\author{S. Nasiri }
\affiliation{Department of Physics, Shahid Beheshti University, G.C., Daneshjou Boulevard, District 1, 19839 Tehran,  Iran}

\author{F. S. Tabatabaei}
\affiliation{School of Astronomy, Institute for Research in Fundamental Sciences (IPM), 19395-5531 Tehran, Iran}

%\collaboration{6}{(AAS Journals Data Editors)}

%\author{Butler Burton}
%\affiliation{}
%\affiliation{}




\begin{abstract}

Magnetic fields can play an important role in the energy balance and formation of gas structures in galaxies. 
However, their dynamical effect on the rotation curve of galaxies is immensely unexplored.
We investigate the dynamical effect of the known magnetic arms of $NGC\;6946$ on its circular gas rotation traced in HI, considering two dark matter mass density models, ISO, and the universal NFW profile. 
We used a three-dimensional model for the magnetic field structure to fit the modeled rotation curve to the observed data via an $\chi$-squared minimization method.
The shape of the HI gas rotation curve is reproduced better including the effect of the magnetic field, especially in the outer part, where the dynamical effect of the magnetic field could become important.
The typical amplitude of the regular magnetic field contribution in the rotation curve is about $ 6 - 14 \;  km s^{-1}$ in the outer gaseous disk of the galaxy $NGC\;6946$. The contribution ratio of the regular magnetic field to the observed circular velocity and to dark matter increases with the galactocentric radius. Its ratio to the observed rotational velocity is about five percent and to dark matter is about 10 percent in the outer regions of the galaxy $NGC\;6946$. 
Therefore, the large-scale magnetic fields cannot be completely ignored in the large-scale dynamics of spiral galaxies, especially in the outer parts of galaxies.

\end{abstract}


\section{Introduction} \label{sec:intro}

\noindent Galaxies contain stars, the interstellar medium (ISM), and dark matter. As one of the main components of the ISM, magnetic fields have a significant energy density and therefore may play an important role in the evolution and dynamics of the host galaxies \citep[e.g.,][]{Beck2007, Tabatabaei2008, beckandHoernes1996}. Magnetic fields couple and interact with gas: They can influence the formation of spiral arms \citep{Kim2002, Gomez2002}, control molecular clouds against gravitational collapse, and decelerate star formation \citep{Heiles2005, Tabatabaei2018}, and help in the onset of galactic outflows \citep[e.g.,][]{Tabatabaei2022}.
The galactic magnetic fields could be considered as a non-negligible dynamical component in the dynamical motion and the rotation of the gaseous distribution of the galactic disk \citep[e.g.,][]{Chyzy2017,  Ruiz-Granados2012} and therefore can reduce dark matter mass needed to explain the rotation curves of galaxies \citep{Battaner2000, Ruiz-Granados2010, Ruiz-Granados2012, Jalocha2012, Sanchez-Salcedo2013}. 
The HI gas rotation may be influenced by the magnetic fields, especially in the outer regions, as the radial decrease of the large-scale magnetic field strength is slower than that of the gravity \citep{Battaner2000, Battaner2007}. \cite{Beck2007} suggests that the slow decrease of the magnetic field energy density in the galaxy $NGC\;6946$ to large radii may influence the gas dynamic in the outer region of the galaxy.\\
Spiral galaxies generally have large-scale magnetic fields with a considerable regular component that has a dominant azimuthal (toroidal) field structure, $B_{\varphi}$. \cite{Beck2004} has presented data on the regular magnetic fields in spiral galaxies. Magnetic forces and the regular component of the magnetic field are high enough to be taken into account in their dynamics and their rotation curve, especially at large radii, in the outermost parts of the galactic disk \citep{ Ruiz-Granados2010, Ruiz-Granados2012, Battaner2007, Battaner2000, Battaner1995, Battaner1992}.
The outermost regions of galactic disks are observed via the $21 cm$ emission line of $HI$ gas, which is sufficiently ionized for the galactic magnetic field lines to be frozen in \citep{Battaner1992}. 
\cite{Nelson1988} studied the dynamical effect of magnetic fields on the rotation curve in the outer gaseous disks of galaxies using a simplified cylindrical model. He obtained rotation velocities higher than the gravitational rotational velocities based on the baryonic matter, caused by the inward force due to the magnetic stress. 
\cite{Battaner1992} proposed magnetism as an alternative explanation for dark matter halo in galaxies. \cite{Battaner1995} also presented a magneto-hydro-dynamically driven rotation curve model with the Magneto-Hydrodynamics (MHD) equation of motion and showed that a magnetic field with a strength of about $1 \mu G$ in the outer regions was able to explain the rotation curve of a spiral galaxy. \cite{Ruiz-Granados2010, Ruiz-Granados2012} claimed that considering the dynamical effects of the large-scale magnetic fields can explain the rising up in the observational HI gas rotation curve for some galaxies. \cite{Ruiz-Granados2010, Ruiz-Granados2012} also concluded that by including the contribution of the regular magnetic fields (the azimuthal component of magnetic field), the shape of the HI rotation curves of M31 and the Milky Way are fitted better, especially in the outer region of the HI disks and therefore a smaller contribution of dark matter is needed in presence of magnetic forces.
For a sample of nearby galaxies, \cite{Tabatabaei2016} found a correlation between the large-scale ordered magnetic field $B_{ord}$ and the rotational velocity of galaxies taken from the flat part of their rotation curves indicating that more rapidly rotating galaxies have stronger magnetic fields than more slowly rotating galaxies.
\cite{Tsiklauri2011} modeled the rotation curve via the Newtonian gravity plus magnetic force, $J \times B$ force, without non-baryonic dark matter, and performed a fit to the Milky Way rotational curve. Their model provided a fair fit to the observed rotation curve of the galaxy for $R \geq 15 kpc$. So they concluded that the effect of $J \times B$ force on the rotation curve of gas for $R \geq 15 kpc$ may be important.\\
Observations of face-on galaxies imply the existence of magnetic spiral arms \citep{Beck2015, beckandHoernes1996, Krause1993IAU}.
Almost all disk galaxies have large-scale spiral magnetic field patterns, even those with little or no evidence of optical spiral arm structures \citep{Nixon2018}. \\
Recently, a successful scale-invariant version of the classical dynamo model has been developed by \cite{Henriksen2017} and \cite{Henriksen2018a}, which subsequently applied for modeling the edge-on spiral galaxy $NGC 4631$ by \cite{Woodfinden2019}. 
By this scale-invariant model, \cite{Henriksen2017} predicted the presence of magnetic spiral arms in the radio halos of galaxies.
The explicit theoretical calculation of the type of scale-invariant gravitational spiral arm associated with the magnetic scale-invariant arm has been presented in \cite{Henriksen2017} to explain theoretically interwoven magnetic and gravitational logarithmic spiral arm in which a gravitational spiral arm may move ahead of the magnetic arm at large radius so that the magnetic arm lies inside the gravitational arm. \\
Recent observation of the magnetic field topology in spiral galaxies which are viewed edge-on shows a plane-parallel magnetic field component in the disk of the galaxy and an X-shaped poloidal magnetic field structure, above the plane in its halo, sometimes accompanied by strong vertical magnetic field components in the near halo above and below the central region of the galactic disk  \citep{Beck2015Rev, Henriksen2016, Henriksen2017, Woodfinden2019, Nixon2018, Stein2019, Krause2009, Krause2015, Wiegert2015, Hanasz2009}.
In flat, rotating galactic disks, the magnetic field is composed of a toroidal component with azimuthal symmetry and spiral structure within the disk and a weaker poloidal component with vertical symmetry (even or odd parity) \citep{Beck2015Rev}. 

\noindent \cite{Tabatabaei2008} studied the magnetic field structure of the Scd galaxy $M 33$ and indicated the presence of a vertical magnetic field component $B_{z}$ in the regular magnetic field $B_{reg}$, by fitting a parameterized model of $B_{reg}$ that includes both horizontal and vertical component to the multi-wavelength radio polarization observations \citep[the observed distributions of the polarization angle at different wavelengths, see also][]{Berkhuijsen1997, Fletcher2004}.\\
By studying the radio continuum emission data of a sample of 35 nearby edge-on galaxies from the continuum halos in nearby galaxies, an EVLA Survey \citep[CHANG-ES,][]{CHANGES} project, \cite{Wiegert2015} showed that a typical spiral galaxy is surrounded by a radio halo of magnetic fields and cosmic rays. The observational radio continuum data have been enhanced by the  CHANG-ES (Continuum Halos in Nearby Galaxies) survey of 35 nearby edge-on galaxies \citep{CHANGESIrwin2019}.
The X-type magnetic fields have been observed for this sample of 35 nearby edge-on galaxies throughout the CHANG-ES survey \citep{CHANGESIrwin2019, Wiegert2015}.

\noindent Magnetic fields are observed to be an important ingredient in disk-halo interactions, in which star formation activity in the disk drives convective vertical hot gas flows upward into the halo, carrying spiral magnetic fields along with the outflowing hot gas motion (vertical magnetic field transport, because of the coupling between the magnetic field and fluid), as predicted in MHD simulations \citep{Norman1989, Hanasz2009}.
Developments in dynamo theory imply that there is a connection between the halo and the galactic disk magnetic fields; the spiral magnetic fields rise into the radio halo, so that magnetic flux is transported from the galactic disk into the halo \citep{Beck2015Rev, Henriksen2016, Henriksen2017, Woodfinden2019, Nixon2018, Stein2019, Krause2009, Krause2015, Wiegert2015, Hanasz2009}.

\noindent \cite{Henriksen2016} presented an analytic model of a magnetized galactic halo around a razor-thin Mestel gravitating disk rotating with uniform velocity to obtain the observed X-shaped topology of the halo magnetic field in the $\rho-z$ plane. The X-shaped structure of the halo magnetic field is predicted to be an intermediate phenomenon, lying between vertical fields (perpendicular to galactic disks, $B_{z}$ component) near the disk and more radial fields at high z, in which the angle of the magnetic field lines depends on the ratio of the vertical to radial components of velocity. \\ 
\cite{Nixon2018} proposed that the action of vertical shear on an initially poloidal magnetic field generates polarization patterns of the observed spiral magnetic fields, i.e. spiral magnetic field patterns nearly parallel in the galactic plane (galaxies seen face-on) and poloidal X-shaped above the plane (galaxies seen edge-on) (see also \cite{Henriksen2016}).
The poloidal magnetic field topology in the upper halo of galaxies can be considered as a linear combination of two different field types, an extension of the thick disk axisymmetric spiral quadrupole magnetic field out to larger heights above the mid-plane and a radially directed dipole magnetic field \citep{Braun2010}.\\
The aim of this paper is to study the dynamical effect of the large-scale magnetic field on the rotation curve of the spiral galaxy $NGC\;6946$  in the presence of the dark matter halo. This spiral galaxy is a regularly rotating galaxy, closely face-on with HI inclination of $35^{\circ}$ \citep{deBlok2008}, and a high radio surface brightness \citep{Sofue1986}.
The observational high-resolution data on the $HI$ gas disk rotation curve of the nearby spiral galaxy $NGC\;6946$ at large radii \citep{deBlok2008} and the recent increased amount and quality of data on linearly polarized radio continuum emission and regular magnetic field observations for this galaxy \citep{Beck2007} motivated us to re-investigate the role of the three dimensional regular magnetic fields in the rotation of gas, and to test whether magnetic fields should be considered as a non-negligible dynamical ingredient.
The observations of polarized radio emission of the gas-rich spiral galaxy $NGC\;6946$ at 6.2 cm wavelength, revealed this galaxy hosts two surprisingly main symmetric magnetic spiral arms, parallel to the adjacent optical spiral arms, located in interarm regions (between the optical spiral arms), with the perfectly aligned regular field. These magnetic spiral arms are more symmetric than the optical spiral arms \citep{beckandHoernes1996, Beck2007, BeckandWielebinski2013, Beck2001, Becketal1996}.\\
This paper is structured in the following order.
In Sec. \ref{ModelB}, we will model the three-dimensional regular magnetic field structure caused by two main magnetic arms to estimate its dynamical influence on the $HI$ gas motion and its contribution to the rotation curve.
We will also map the distribution of planar component of the regular galactic magnetic field strength in the galactic plane with some models: the axisymmetric spiral field ($m = 0$ mode, ASS), the concentric-ring model, the bisymmetric spiral field ($m = 1$ mode, BSS), and quadri-symmetric spiral filed ($m = 2$ mode, QSS). We will also generate the map of the planar component of the regular galactic magnetic field strength for the superposition of two azimuthal dynamo modes $m = 0$ and $m = 1$ and the superposition of two azimuthal dynamo modes $m = 1$ and $m = 2$ with the same amplitudes.
Then in Sect. \ref{Modeling the Rotation Curve}, considering the three-dimensional regular magnetic field, we will model the rotation curve by means of the contribution from different components to describe the observed rotation curve of $NGC\;6946$, for two dark matter halo models, ISO and NFW profile.
In this section, to investigate whether the large-scale magnetic fields should be considered as a non-negligible dynamical ingredient to the $HI$ gas rotation, we will present the best fits solution to the observed rotation curve in two different cases, i.e., with and without including the contribution of the regular magnetic field to the circular velocity, with two dark matter mass density models, ISO and the universal NFW profile.
We also obtain the contribution of the regular magnetic field caused by the two main inner magnetic arms to the $NGC\;6946$ rotation curve and will construct the map of the modeled regular magnetic field strength for the spiral galaxy $NGC\;6946$.
Finally, in Sect. \ref{Dis} and Sect. \ref{Conclusion}, we will discuss and summarize the concluding remarks, respectively.

\section{Modeliing the Magnetic Field Structure in $NGC 6946$}\label{ModelB}

\noindent The magnetic fields in spiral galaxies are an important component, with two of the known main characteristics of their basic three-dimensional topology; 1: the fields in nearly face-on spiral galaxies are observed to follow the spiral pattern traced in the optical morphology, and 2: field distributions which are seen to extend into the halo regions (spiral magnetic fields rising into the halo) and have a characteristic X-shaped morphology \citep{Henriksen2016, Henriksen2017, Heesen2009, Braun2010}.
Therefore we consider the regular magnetic configuration in three dimensions with spiral magnetic field patterns nearly parallel in the galactic plane (galaxies seen face-on) and poloidal X-shaped above the plane (galaxies seen edge-on) \citep{Henriksen2016, Nixon2018}.
\noindent To model three dimensional structure of the regular galactic magnetic field, we first model the regular magnetic field in the galactic plane, and then we modify the field configuration to three dimensions by adding the vertical component of the regular magnetic field.  

\subsection{Models of the planar regular galactic magnetic field structures in the galactic plane and mapping the regular galactic magnetic field strength}

\noindent The observational results of the regular magnetic field of spiral galaxies suggest that the magnetic field geometry is dominated by an in-plane magnetic field component \citep{Braun2010}.

\noindent Morphology of the regular magnetic fields and the radio-polarization pattern in the galactic plane can be described based on its dominant mode, with some models: the axisymmetric spiral field (azimuthal mode $m = 0$, ASS), the concentric-ring model, the bisymmetric spiral field ($m = 1$ mode, BSS), and quadri-symmetric spiral filed ($m = 2$ mode, QSS) or higher harmonics \citep{vallee1991, vallee1992, Krause1990IAUS, Han1994, Han2006, poezd1993, Sofue1983, Sofue1986, Indrani1999, Jansson2009, RK1989, Ruiz-Granados, Arshakian2007, Braun2010}. \\

%\bigskip
%\begin{itemize}
%\item {\bf Axisymmetric Model}
%\end{itemize}
{\bf Axisymmetric Model}\\

\noindent One of the simplest descriptions of the galactic magnetic field is the axisymmetric model (ASS, $m = 0$) \citep{vallee1991, poezd1993, Ruiz-Granados}. The $m = 0$ azimuthal dynamo mode generates an axisymmetric spiral field that points in the same direction all along a circle around the galaxy$ ^{,} $s center (same pitch angle).
By assuming the regular component of the magnetic field to be mainly parallel to the galactic plane ($B_{z} = 0$), both the azimuthal and radial components of the regular magnetic field contribute to the $HI$ gas disk rotation curve of the spiral galaxies. The azimuthal component of the regular magnetic field contributes to an inward-directed Lorentz force, but the radial component is not negligible in galaxies. Since the magnetic fields have zero divergences $\triangledown. B = 0$ (the solenoidality condition), if , $B_{z} = 0$, for the axisymmetric spiral field (azimuthal mode $m = 0$, ASS), one can obtain $\partial (\rho B_{\rho})/\partial \rho = 0$, and then $B_{\rho} \propto 1 / \rho$, thus giving a singularity on the axis! \citep{MossandSokoloff2019}. The singularity problem at the origin can be avoided by considering the components of  the regular magnetic field for the ASS model as follow \citep{Ruiz-Granados2012}:  

\begin{align}\label{ASS}
B_{\rho} =& B_{0}(\rho) \sin(p) \nonumber\\
B_{\phi} =& B_{0}(\rho) \cos(p)
\end{align}
with
\begin{equation}\label{r1}
B_{0}(\rho) = \frac{B_{1}}{1 + \frac{\rho}{\rho1}}
\end{equation}
where $p$ is the pitch angle and $B_{0}(\rho)$ is the regular magnetic field strength as a function of the radial distance $\rho$. $B_{1}$ and $\rho_{1}$ are constant. The radial factor $\rho_{1}$ represents the characteristic scale at which $B_{0}(\rho)$ decreases to half its value at the galactic center. \\

%\bigskip
%\begin{itemize}
%\item {\bf Concentric Ring Model}
%\end{itemize}
{\bf Concentric Ring Model}\\

\noindent This model which presents reversals of the magnetic field as a function of the radius has only an azimuthal component, $B_{\phi}$ \citep{Indrani1999}. 
\begin{equation}
B_{\phi} =B_{0} \sin\left[ \dfrac{\pi}{\omega}\left\lbrace  \rho- (R_{0} - d_{\rho})\right\rbrace  \right] , B_{\rho} = 0
\end{equation}
where $\omega$ is the space between the reversals, $d$ is the distance of the first reversal and $B$ is the amplitude of the magnetic field strength.\\

%\bigskip
%\begin{itemize}
%\item {\bf Bisymmetric Spiral Model}
%\end{itemize}
{\bf Bisymmetric Spiral Model}\\

\noindent \cite{Han1994} and \cite{Indrani1999} have modeled the regular magnetic field in the disk of our galaxy in this region as a bisymmetric spiral configuration.
The structure of this model for the galactic magnetic field, in the polar coordinate system, can be expressed as: 

\begin{align}
B_{\rho} =& B_{0}(\rho) \cos(\varphi - \beta \ln(\frac{\rho}{\rho_{0}})) \sin(p) \; , \nonumber\\
B_{\varphi} =& B_{0}(\rho) \cos(\varphi - \beta \ln(\frac{\rho}{\rho_{0}})) \cos(p)
\end{align}
where $\beta $ is a constant, $\beta = 1/\tan(p)$, and $p$ is the pitch angle.
The regular magnetic field of the Milky way is assumed to be mainly parallel to the galactic plane \cite{Han1994}.
Therefore, at the point $(\rho, \varphi)$, the magnetic field strength is:
\begin{equation}\label{1-}
B(\rho, \varphi) = B_{0}(\rho) \cos(\varphi - \beta \ln(\frac{\rho}{\rho_{0}})).
\end{equation}
At the point $(\rho_{0}, \phi = 0 )$, the field reaches the first maximum \citep{Han1994, Jansson2009}.\\
There is another possible family of bisymmetric models, which corresponds to a positive sign inside the parenthesis instead of the negative sign, which was proposed by \citep{Jansson2009}:
\begin{equation}\label{1+}
B(\rho, \varphi) = B_{0}(\rho) \cos(\varphi + \beta \ln(\frac{\rho}{\rho_{0}})).
\end{equation}
Positive values of magnetic field strength indicate clockwise magnetic field lines, while negative values of magnetic field strength indicate counterclockwise field lines \citep{vallee1991}. \\

%\bigskip
%\begin{itemize}
%\item {\bf Quadri-symmetric Spiral Model}
%\end{itemize}
{\bf Quadri-symmetric Spiral Model}\\

\begin{figure*} 
\subfigure[]{ \includegraphics[width=1\columnwidth]{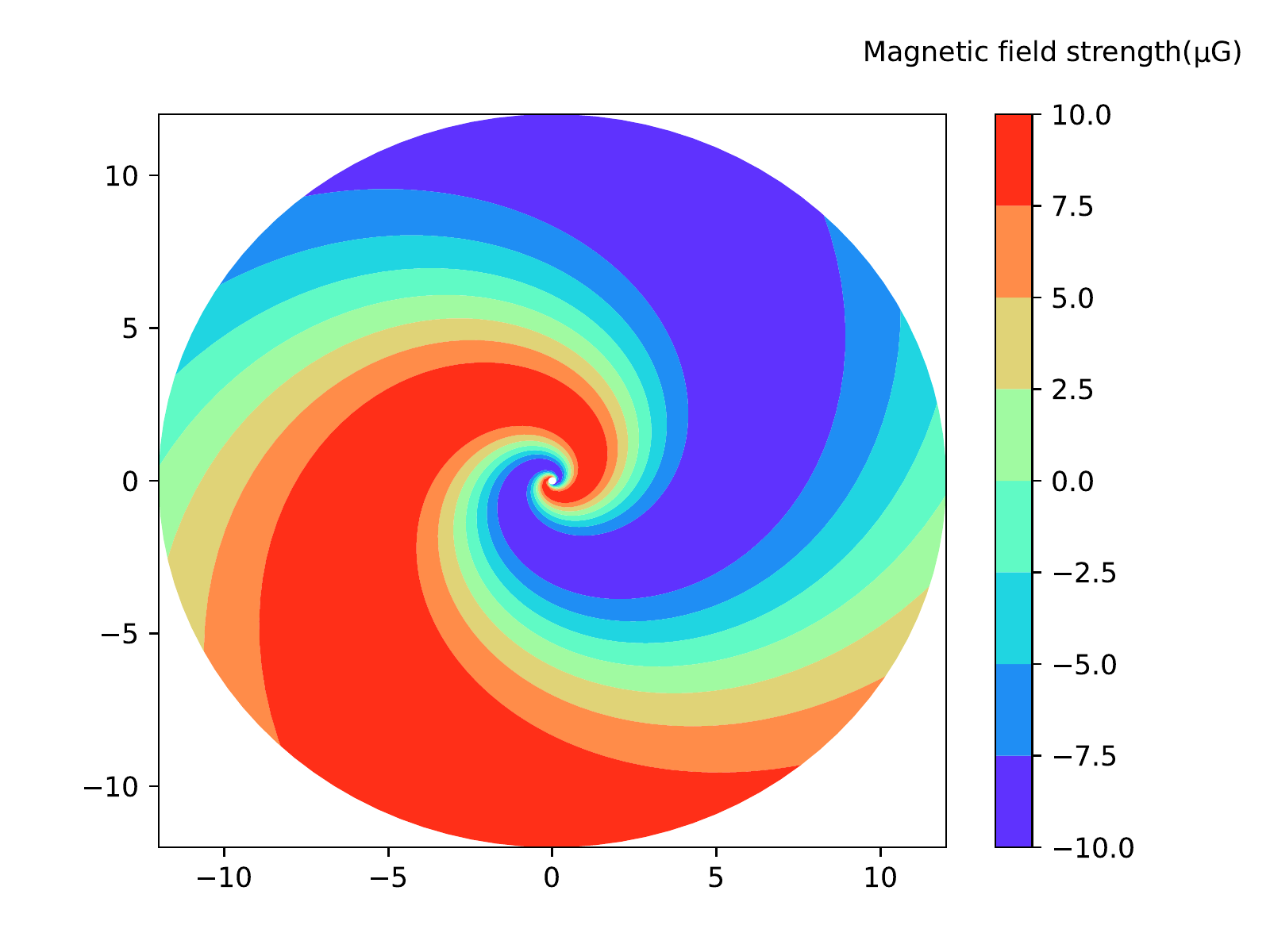}}\quad
   \subfigure[]{ \includegraphics[width=1\columnwidth]{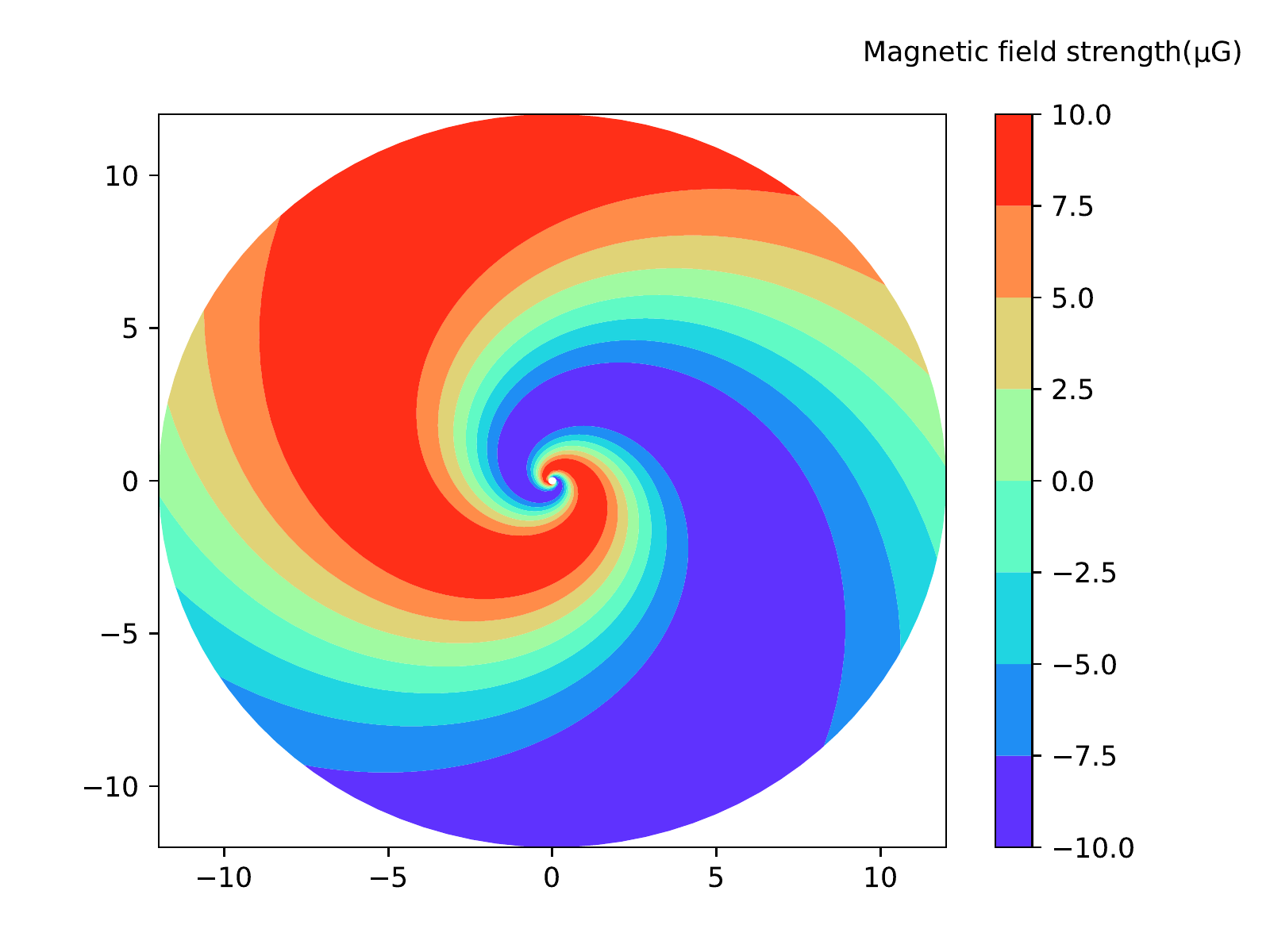}}
    \caption{Map (color-coded image) of the planar component of the modeled regular galactic magnetic field strength in the galactic disk plane (in the $X-Y$ plane), constructed for the regular field mode m = 1, Bisymmetric Spiral Model with the field strength amplitude $B_{0} = 10$  $\mu G$, (a): corresponds to a negative sign inside the parenthesis of Eq. \ref{1-} and (b): corresponds to a positive sign inside the parenthesis of Eq. \ref{1+}.  
Positive values of magnetic field strength indicate clockwise magnetic field lines, while negative values of magnetic field strength indicate counterclockwise field lines. The BSS mode has two reversals. The $m = 1$ azimuthal dynamo mode generates two magnetic spiral arms with alternating field directions.}
\label{m=1}
\end{figure*}

\begin{figure*} 
\subfigure[]{ \includegraphics[width=1\columnwidth]{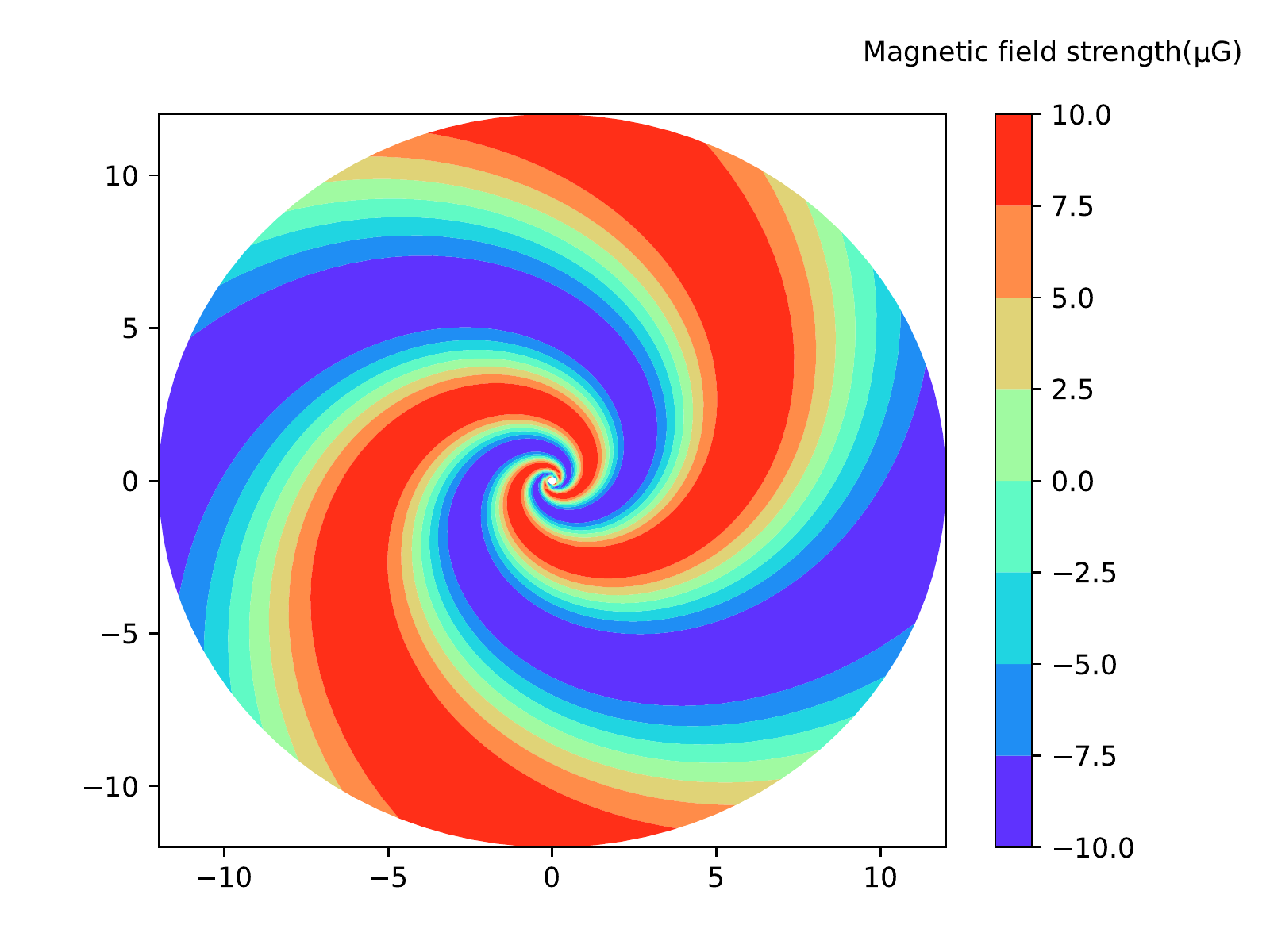}}\quad
   \subfigure[]{ \includegraphics[width=1\columnwidth]{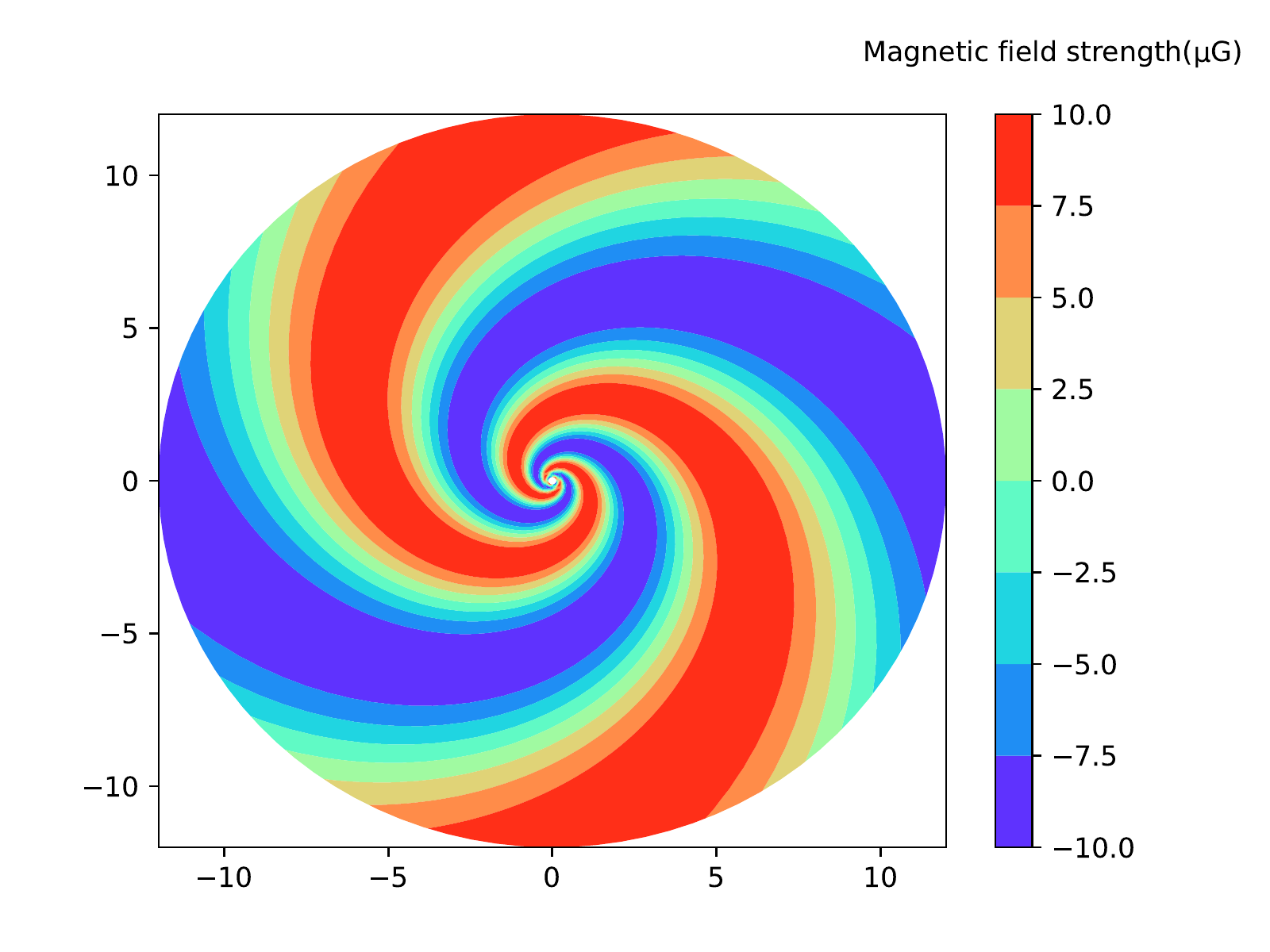}}
    \caption{Map (color-coded image) of the planar component of the modeled regular galactic magnetic field strength in the galactic disk plane (in the $X-Y$ plane), constructed for the regular field mode $m = 2$, Quadri-symmetric Spiral Model with the field strength amplitude $B_{0} = 10$ $\mu G$, (a): corresponds to a negative sign inside the parenthesis of Eq. \ref{2-+} and (b): corresponds to a positive sign inside the parenthesis of Eq. \ref{2-+}. Positive values of magnetic field strength indicate clockwise magnetic field lines, while negative values of magnetic field strength indicate counterclockwise field lines. The QSS mode has four reversals along the azimuthal angle. The $m = 2$ azimuthal dynamo mode generates four magnetic spiral arms with alternating field directions.}
\label{m=2}
\end{figure*}

\noindent Explaining the morphology of the regular magnetic fields in some galaxies needs a dominant higher mode, $m=2$, which is classified as quadri-symmetric spirals (QSS). 
The $m = 2$ azimuthal dynamo mode generates four magnetic spiral arms with alternating field directions. 
The QSS mode has four reversals along the azimuthal angle, while the BSS mode has two reversals \citep{vallee1992, Arshakian2007}.
The strength of the regular magnetic field for this model in cylindrical coordinates is defined as

\begin{equation}\label{2-+}
B(\rho, \varphi) = B_{0}(\rho) \cos \left[ 2 \left(\varphi \mp \beta \ln(\frac{\rho}{\rho_{0}})\right) \right]. 
\end{equation}
By assuming the regular component of the magnetic field to be mainly parallel to the galactic plane, the structure of this model for the galactic magnetic field can be expressed as: 

\begin{align}
B_{\rho} =& B_{0}(\rho) \cos \left[ 2 \left(\varphi \mp \beta \ln(\frac{\rho}{\rho_{0}})\right) \right]  \sin(p) \; , \nonumber\\
B_{\varphi} =& B_{0}(\rho) \cos \left[ 2 \left(\varphi \mp \beta \ln(\frac{\rho}{\rho_{0}})\right) \right]  \cos(p).
\end{align}
We mapped the distribution of the regular galactic magnetic field strength in the galactic plane for the axisymmetric model (ASS, $m = 0$ mode) and concentric-ring model in Fig.\;(\ref{m=0, Concentric ring}), bisymmetric spiral model (BSS, $m = 1$ mode) in Fig.\;(\ref{m=1}), and quadri-symmetric spiral model (QSS, $m = 2$ mode) in Fig.\;(\ref{m=2}). 
We also generated the map of the regular galactic magnetic field strength for the superposition of two azimuthal dynamo modes $m = 0$ and $m = 1$ and the superposition of two azimuthal dynamo modes $m = 1$ and $m = 2$ with the same amplitudes in Fig.\;(\ref{m=0,1,2}).

\begin{figure*} 
\subfigure[]{ \includegraphics[width=1\columnwidth]{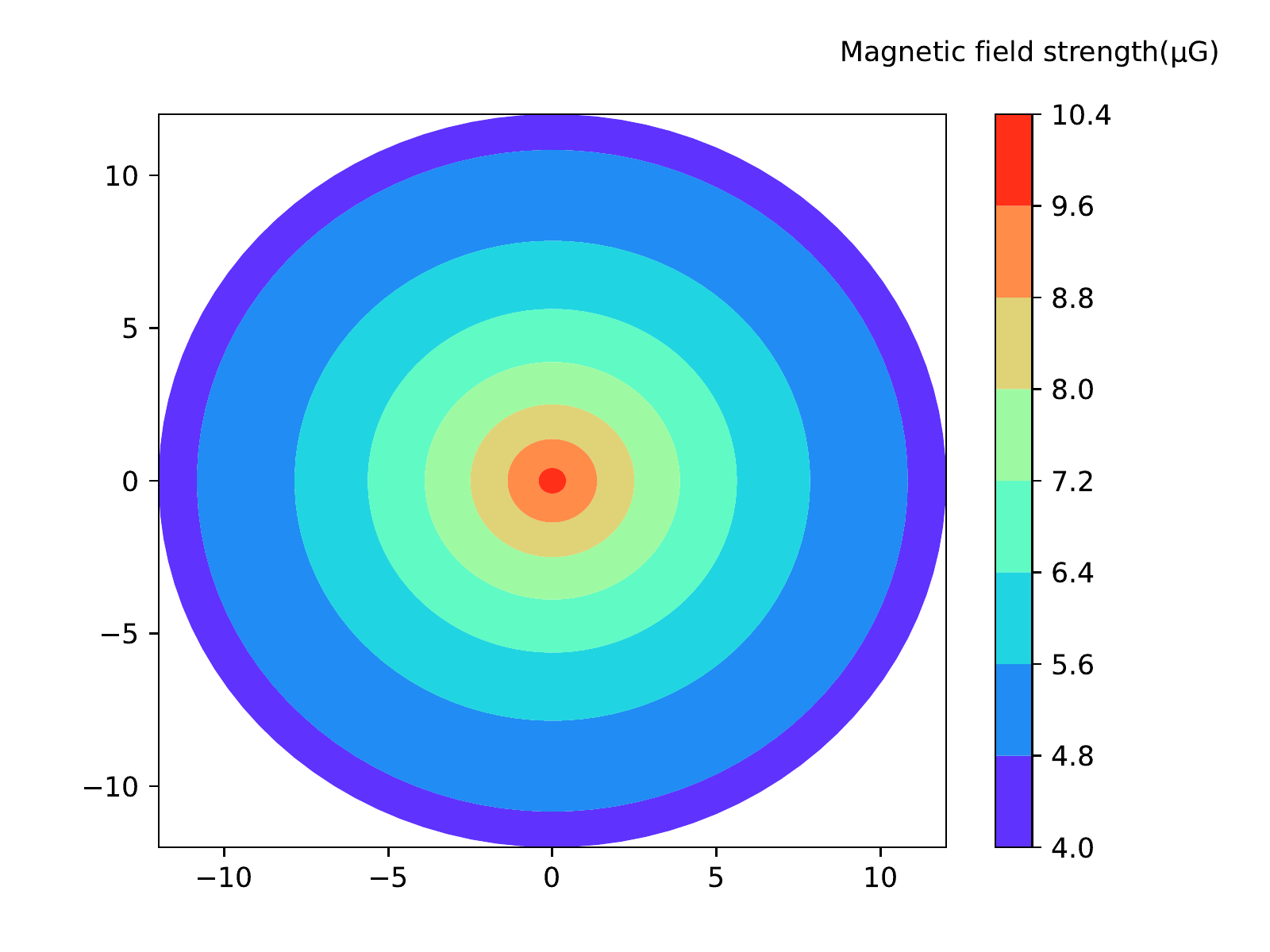}}\quad
   \subfigure[]{ \includegraphics[width=1\columnwidth]{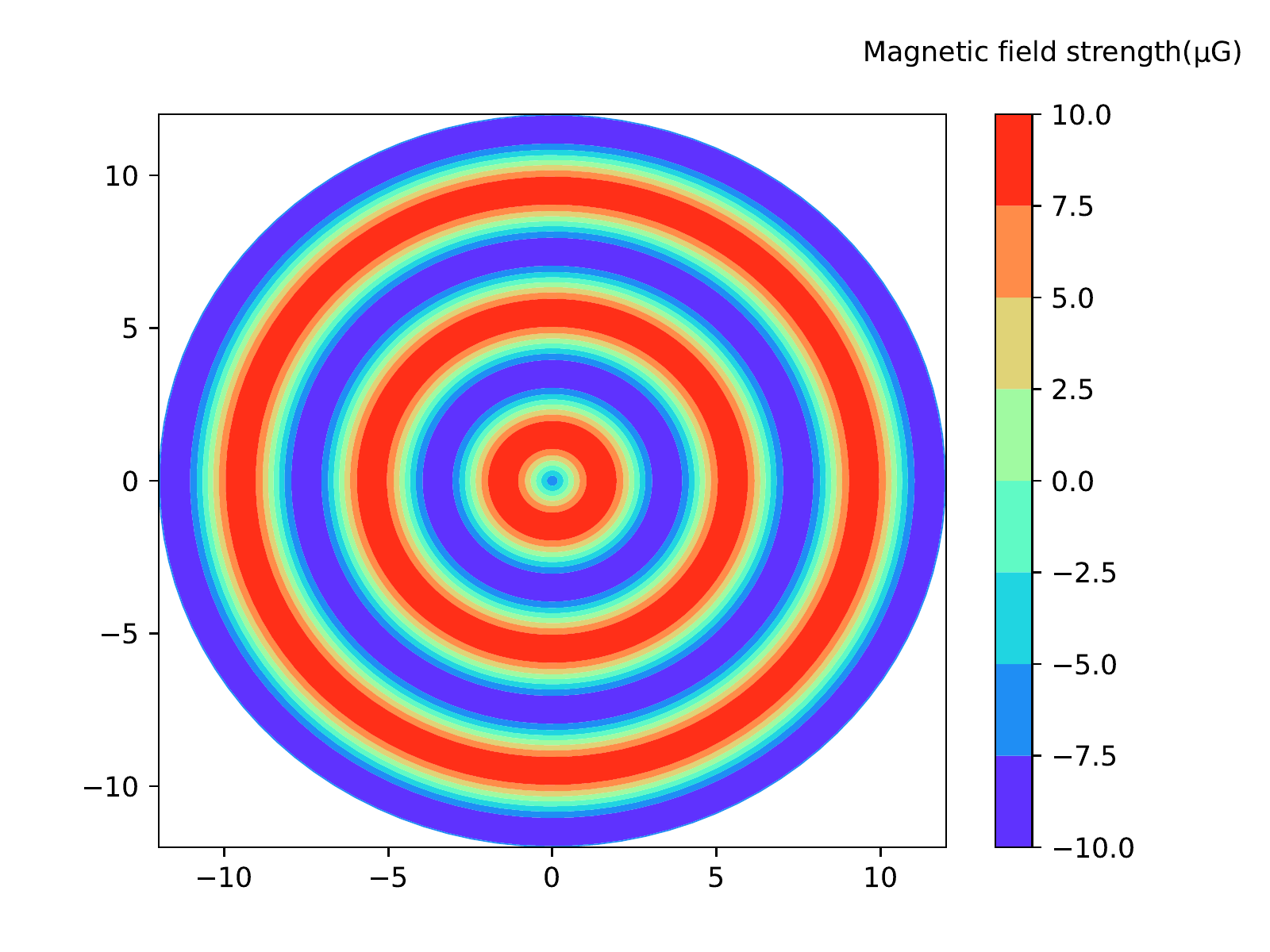}}
    \caption{Map (color-coded image) of the planar component of the modeled regular galactic magnetic field strength in the galactic disk plane (in the $X-Y$ plane) for (a): the regular field mode $m = 0$, Axisymmetric Model (ASS), and (b): for Concentric Ring Model. The $m = 0$ azimuthal dynamo mode generates an axisymmetric spiral field that points in the same direction all along a circle around the galaxy’s center (same pitch angle), while the concentric ring model describes reversals of the magnetic field as a function of  radius. Positive values of magnetic field strength indicate clockwise magnetic field lines, while negative values of magnetic field strength indicate counterclockwise field lines.}
\label{m=0, Concentric ring}
\end{figure*}

\begin{figure*} 
\subfigure[]{ \includegraphics[width=1\columnwidth]{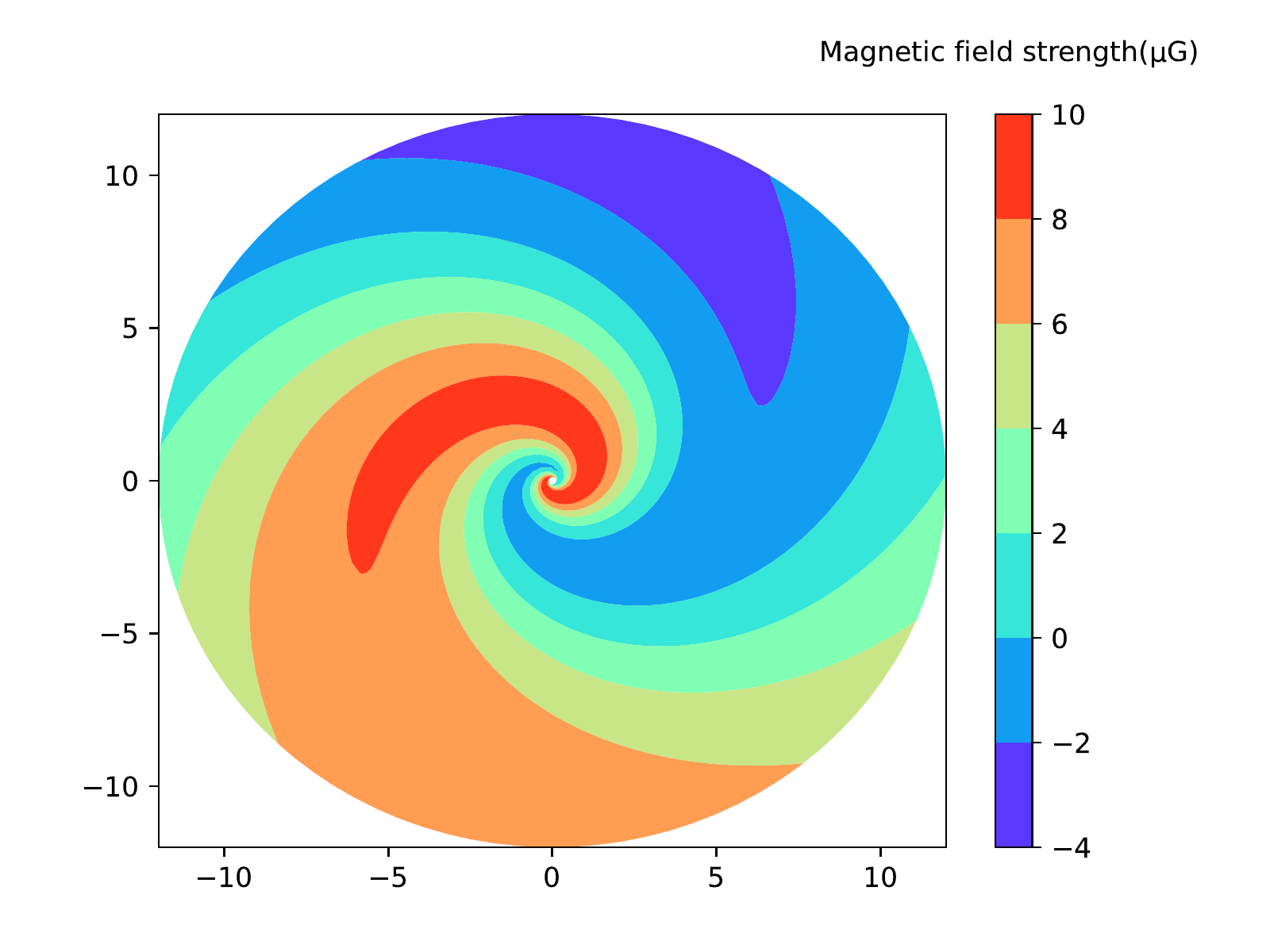}}\quad
   \subfigure[]{ \includegraphics[width=1\columnwidth]{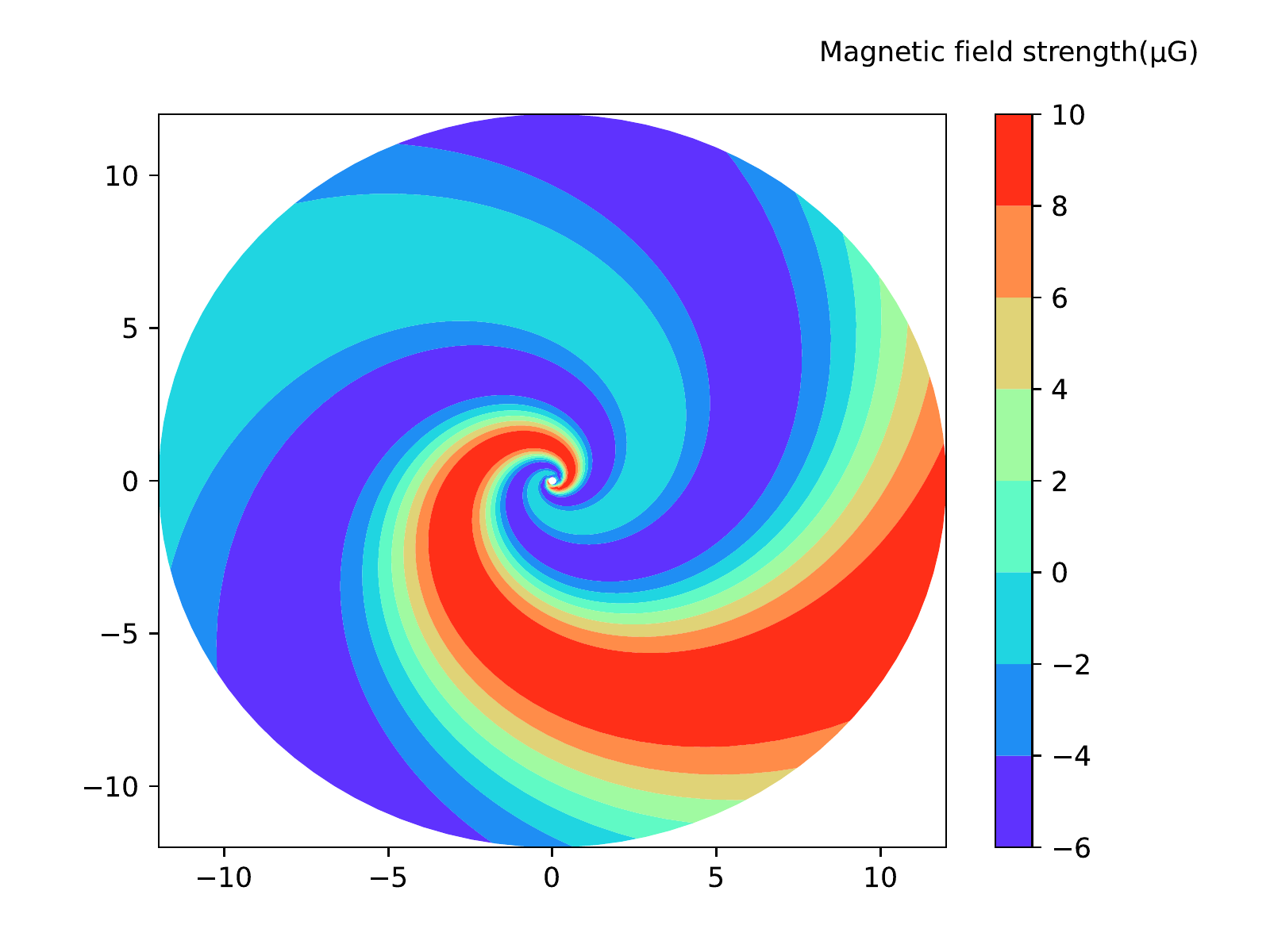}}
    \caption{Map (color-coded image) of the planar component of the modeled regular galactic magnetic field strength in the galactic disk plane (in the $X-Y$ plane), (a): constructed with the superposition of two azimuthal dynamo modes $m = 0$ and $m = 1$ with the same amplitudes, and (b): constructed with the superposition of two azimuthal dynamo modes $m = 1$ and $m = 2$ with the same amplitudes. 
The superposition of the azimuthal modes $m = 0$ and $m = 1$ with the same peak amplitudes, leads to only one dominant magnetic spiral arm, while the superposition of the azimuthal modes $m = 1$ and $m = 2$ with the same peak amplitudes, lead to one main magnetic spiral arm and two secondary magnetic spiral arms. Positive values of magnetic field strength indicate clockwise magnetic field lines, while negative values of magnetic field strength indicate counterclockwise field lines. The magnetic field lines of the main magnetic arm have a clockwise direction, while those of the two secondary magnetic spiral arms have a counterclockwise direction.}
\label{m=0,1,2}
\end{figure*}

\bigskip

\subsection{Three-dimensional Magnetic Field Topology}

\noindent In the first detection of a large-scale magnetic field in a Sa galaxy ($M104$) in the radio range, \cite{Krause2006} found that however, the magnetic field is predominantly plane parallel (a mainly toroidal magnetic field configuration parallel to the disk of the galaxy), there are deviations from the alignment of magnetic field line with the disk, which is seen mainly above and below the galactic disk, an indication of vertical magnetic field components at larger $z$-distances from the disk. Therefore the large-scale magnetic
field can be considered as a superposition of a galactic disk parallel and a vertical component.

\noindent To simulate the regular magnetic field structure in three dimensions, one can extend the planar magnetic field topology with dominant  azimuthal (toroidal) mode(s) by adding an out-of-plane poloidal field component taken from a linear combination of dipole and quadrupole magnetic field topologies \citep{Braun2010, Long2007}.

\begin{figure*}\label{DQ}
\subfigure[]{ \includegraphics[width=0.630\columnwidth]{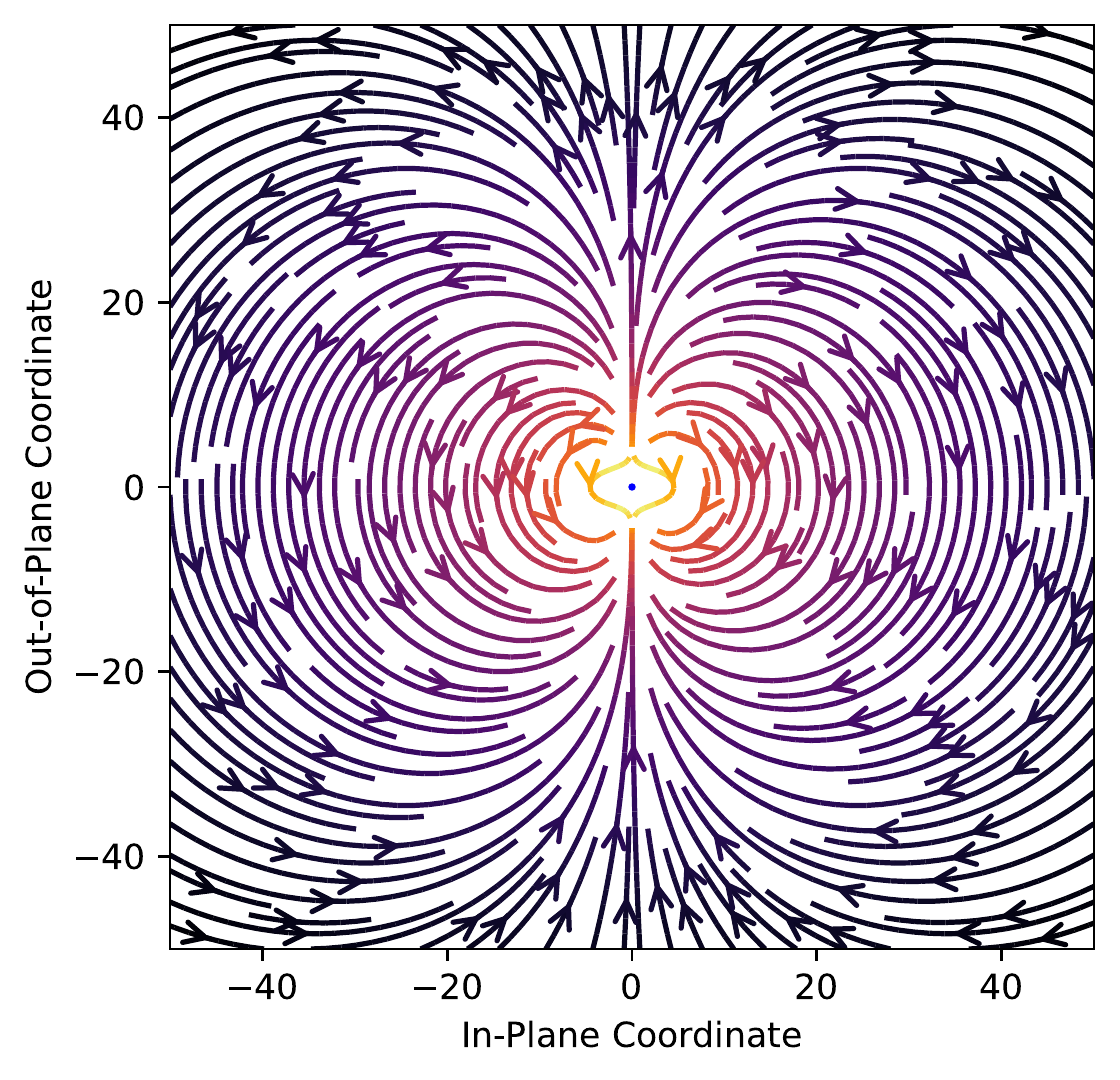}}\quad
   \subfigure[]{ \includegraphics[width=0.630\columnwidth]{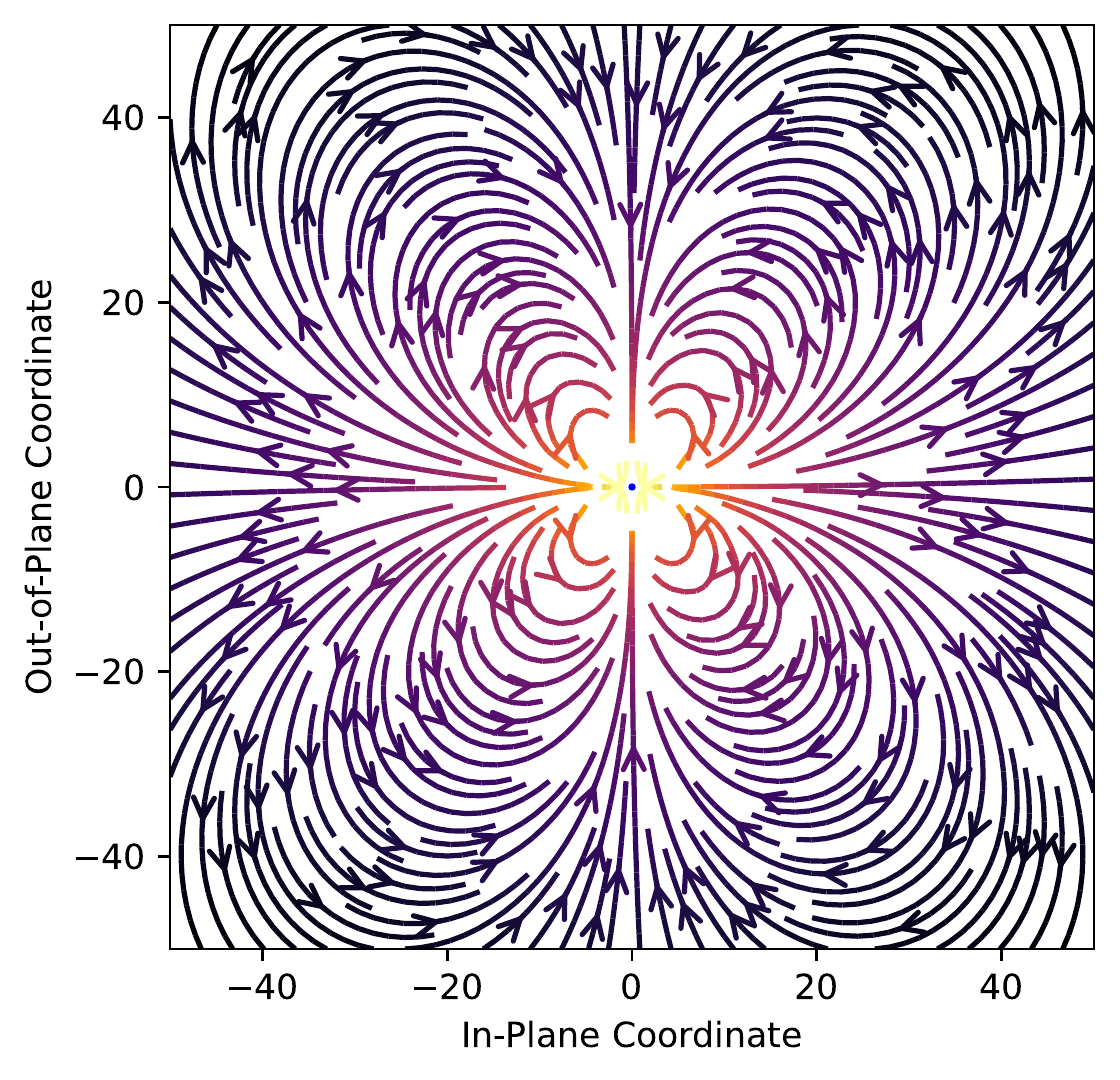}}\quad
      \subfigure[]{ \includegraphics[width=0.630\columnwidth]{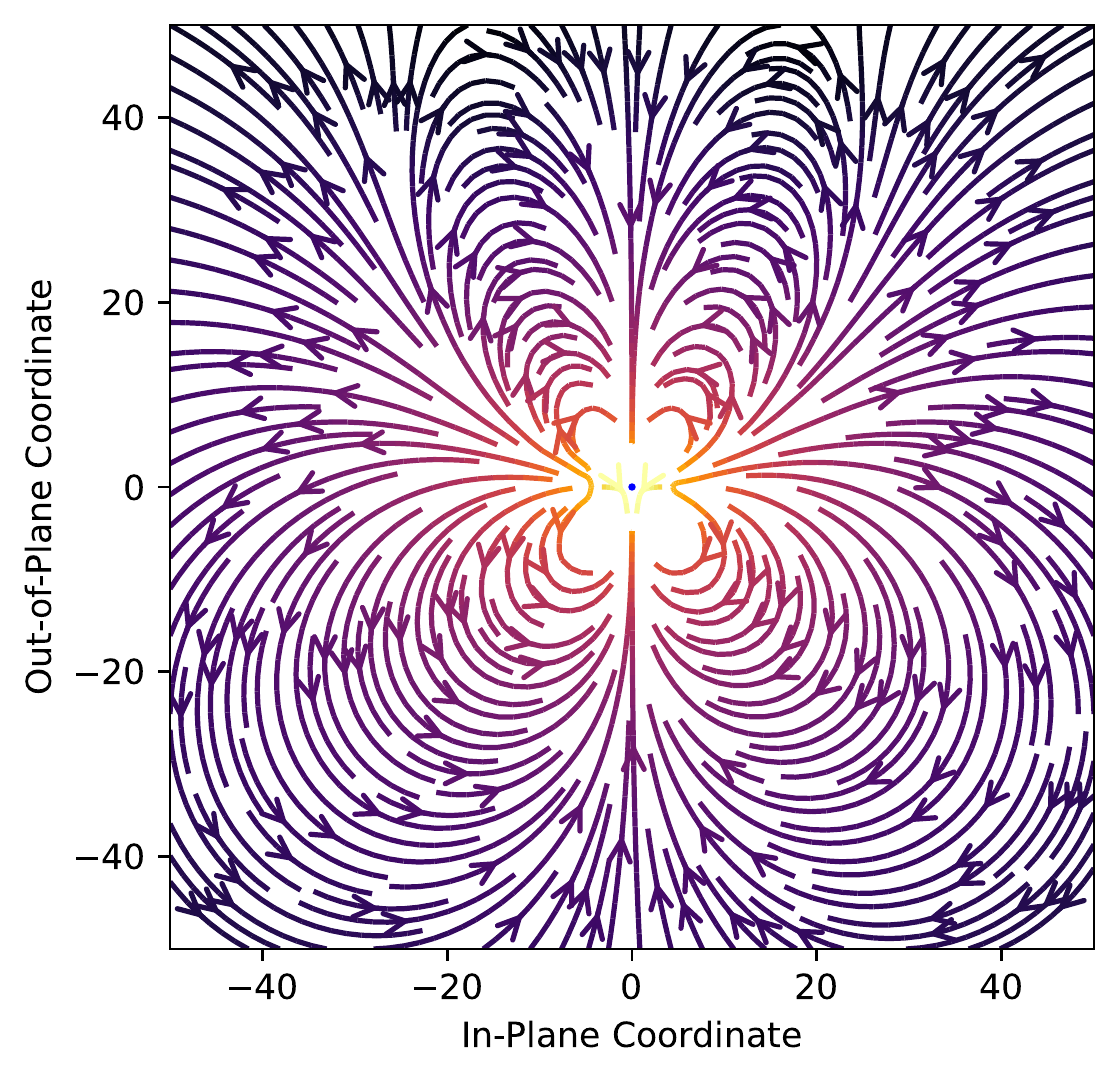}}\\
    \caption{Depiction of the poloidal out-of-plane magnetic field topology that illustrates magnetic field line of (a) a pure dipole, (b) a pure quadrupole, and (c) a 1:100 combination of dipole and quadrupole. A planar logarithmic magnetic spiral is modified by the local orientation of a dipole or quadrupole field or their linear combination, which is symmetric around the rotation axis.}
\label{pole}
\end{figure*}

\noindent In the spherical coordinate system, dipole magnetic field configuration is given by: 
\begin{align}\label{quadrupole}
B_{r} = & \frac{2D}{ r^{3}} \cos(\theta)\\
B_{\theta} =& \frac{D}{r^{3}} \sin(\theta)
\end{align}
in which $D$ is the dipole moment.
The perpendicular component of the poloidal magnetic field with a quadrupole moment, Q, in the spherical coordinate system, $
(r , \theta , \varphi)$ is given by \citep{Long2007, Braun2010}:
\begin{align}
B_{r} = & \frac{3Q}{4 r^{4}} (3 \cos^{2}(\theta) - 1) \label{quadrupoler}\\
B_{\theta} =& \frac{3Q}{2 r^{4}} \cos(\theta) \sin(\theta) \label{quadrupoletheta}
\end{align}
\noindent The poloidal magnetic field topologies in the plane $\rho - z$ have been shown in Fig. \ref{DQ} for pure dipole (a), quadrupole (b), and a combination of dipole plus quadrupole configuration (c). A planar logarithmic magnetic spiral is modified by the local orientation of a dipole or quadrupole field or their linear combination, which is symmetric around the rotation axis.

\noindent Magnetic field components within and out of the plane of the galaxy, in a cylindrical coordinate system, $(\rho, \varphi, z)$ are given by:
\begin{align}
B_{\rho} = & B_{r} \sin(\theta) + B_{\theta}\cos(\theta)\label{Brho}\\
B_{z} = & B_{r} \cos(\theta) - B_{\theta}\sin(\theta)\label{Bz}
\end{align}
\noindent in which the total magnetic field strength, $B_{0}$ and the local orientation angle $\psi_{z}$ are obtained as follow:
\begin{align}\label{}
B_{0} = & \sqrt{B_{z}^{2} + B_{\rho}^{2} }\\
\psi_{z} = & \tan^{-1}(\frac{B_{z}}{B_{\rho}}) \label{ciz}
\end{align}

\noindent Transform from spherical coordinates $ (r, \theta, \varphi) $ to cylindrical coordinates $ (\rho, \varphi, z) $:
\begin{align}
r = & \sqrt{\rho^{2} + z^{2}}\label{SphtoCyl1}\\
\theta = & \arctan (\frac{\rho}{z})\label{SphtoCyl2}\\
\sin(\theta) = & \frac{\rho}{\sqrt{\rho^{2} + z^{2}}}\label{SphtoCyl3}\\
\cos(\theta) = & \frac{z}{\sqrt{\rho^{2} + z^{2}}}\label{SphtoCyl4}
\end{align}

\noindent Like other spiral galaxies, $NGC\;6946$ has spiral magnetic patterns in the disk combined with a mainly quadrupolar poloidal magnetic field component that becomes dominant at large vertical distances from the disk \citep{Braun2010, Heald2012, Henriksen2016, 
Soida2011}.

\noindent By inserting equations \ref{SphtoCyl1} - \ref{SphtoCyl4} in equations \ref{quadrupoler} and \ref{quadrupoletheta}, one can obtain the perpendicular component of the poloidal magnetic field with a quadrupole moment, Q, in the cylindrical coordinate system, as follow:
\begin{align}
B_{r} = &  \frac{3 Q}{4 (\rho^{2} + z^{2})^{2}}\left[ \frac{3 z^{2}}{(\rho^{2} + z^{2})} - 1 \right] \label{BrA1}\\
B_{\theta} = &  \frac{3 Q z \rho}{2 (\rho^{2} + z^{2})^{3}} \label{BthetaA1}
\end{align}
\noindent By inserting equations \ref{BrA1} and \ref{BthetaA1} in equations \ref{Brho} and \ref{Bz}, one rewrite magnetic field components within and out of the plane of galaxy as follow:
\begin{align}
B_{\rho} = &  \frac{3 Q \rho}{4 (\rho^{2} + z^{2})^{\frac{5}{2}}}\left[ \frac{3 z^{2}}{(\rho^{2} + z^{2})} - 1 \right] +   \frac {3 Q z^{2} \rho}{2 (\rho^{2} + z^{2})^{\frac{7}{2}}}\label{down} \\
B_{z} = &  \frac{3 Q z}{4 (\rho^{2} + z^{2})^{\frac{5}{2}}} \left[ \frac{3 z^{2}}{(\rho^{2} + z^{2})} - 1 \right] -  \frac {3 Q z \rho^{2}}{2 (\rho^{2} + z^{2})^{\frac{7}{2}}} \label{up}
\end{align}

\noindent We extend the planar logarithmic spiral magnetic field topology with the addition of an out-of-plane component with quadrupole field \citep{Heald2012} that is symmetric about the rotation axis.
Combining the planar (with azimuthal mode number, m = 1, 2, ...) and out-of-plane geometries, three components of the modified magnetic field in the cylindrical coordinate system can be written as:
\begin{align}\label{}
B_{\rho} = &  B_{0} \cos \left[ m \left(\varphi \mp \beta \ln(\frac{\rho}{\rho_{0}})\right) \right] \sin(p) \cos(\psi_{z}) \nonumber\\
B_{\varphi} = & B_{0} \sin \left[ m \left(\varphi \mp \beta \ln(\frac{\rho}{\rho_{0}})\right) \right] \cos(p) \cos(\psi_{z}) \nonumber\\
B_{z} = & B_{0} \sin(\psi_{z})
\end{align}
\noindent For the axisymmetric model with azimuthal mode $m = 0$, the three-dimensional magnetic fields will be modified as:
\begin{align}\label{ASSModified}
B_{\rho} =& B_{0}(\rho) \sin(p) \cos(\psi_{z}) \nonumber\\
B_{\phi} =& B_{0}(\rho) \cos(p) \cos(\psi_{z}) \nonumber\\
B_{z} = & B_{0}(\rho) \sin(\psi_{z})
\end{align}

\subsection{Regular Magnetic Field Contribution $(\text{Influence of the Magnetic Field on the Gas Rotation})$}

\noindent The presence of a regular magnetic field could affect the dynamical motion of $HI$ gas in the galactic disk by the radial component of the magnetic force \citep{Ruiz-Granados2012, Chyzy2017, Battaner1992, Battaner1995, Battaner2000, Battaner2007}.

\noindent We are interested in the regular component of the galactic disk magnetic field to estimate the contribution of the dynamical effects of magnetic forces to the $HI$ gas rotation. 

\noindent By considering the magnetic forces, under the conditions leading to cosmic MHD, the equation of motion for the fluid element can be written as:
\begin{equation}\label{eqyek}
\varrho \frac{\partial v}{\partial t}+\varrho v.\nabla v+\nabla P = n F+\frac{1}{4\pi}B.\nabla B-\nabla\left(\frac{B^{2}}{8\pi}\right)
\end{equation}
\noindent where $\varrho$ is the gas density, $v$ is the velocity of the fluid, $P$ is the pressure, $n$ is the number density, $F$ is the total force due to gravity, and $B$ is the magnetic field. 
We assume the simplest form of MHD, Ideal MHD, assumes that the fluid has infinite conductivity. 
By assuming axisymmetry and pure rotation:
\begin{equation}
v = (v_{\rho},v_{\phi},v_{z}) =(0,v_{c}, 0).
\end{equation}
So the radial component of the equation of motion in cylindrical geometry is
\begin{equation}
\varrho \left(-\frac{d\Phi(\rho)}{d\rho}+\frac{v_{c}^{2}}{\rho}\right)-\frac{d P}{d\rho}-F_{\rho}^{mag}=0
\end{equation}
where $\Phi(\rho)$ is the gravitational potential, $F^{mag}_{\rho}$ is the radial component of the magnetic force ($ F^{mag}_{\rho} = (J \times B)_{\rho}$), and $P$ is the pressure of
the fluid. 
Following treatment by \citep{Ruiz-Granados2010, Ruiz-Granados2012}, we assume that pressure gradients in the radial
direction to be negligible \citep{Battaner2000}.

\noindent From the equation of motion in fluid dynamics, the contribution of the magnetic field to the circular velocity due to the magnetic force is given by:

\begin{equation}\label{VM}
\begin{split}
v_{mag}^{2}=&\frac{1}{4\pi \varrho} \left[\partial_{\rho}(\rho B_{\varphi}) - \partial_{\varphi}(B_{\rho})\right] B_{\varphi}\\
& - \frac{1}{4\pi \varrho} \left[ \partial_{z}( B_{\rho}) - \partial_{\rho}( B_{z}) \right] \rho  B_{z},
\end{split}
\end{equation}

\noindent The magnetic field is divergence-free:
\begin{equation}
\nabla \cdot B = 0.
\end{equation}
Hence
\begin{equation}
\dfrac{\rho}{4 \pi \varrho} B_{\rho}\nabla \cdot B = 0.
\end{equation}
\noindent Therefore, we can rewrite the first term in Eq. \;(\ref{VM}) as:
\begin{equation}
\begin{split}
\left[\partial_{\rho}(\rho B_{\varphi}) - \partial_{\varphi}(B_{\rho})\right] B_{\varphi} & = \left[\partial_{\rho}(\rho B_{\varphi}) - \partial_{\varphi}(B_{\rho})\right] B_{\varphi} \\
& - \dfrac{\rho}{4 \pi \varrho} B_{\rho}\nabla \cdot B.
\end{split}
\end{equation}
Then
\begin{equation}
\begin{split}
& \left[\partial_{\rho}(\rho B_{\varphi}) - \partial_{\varphi}(B_{\rho})\right] B_{\varphi}   =  \\
& \left[ B_{\varphi} \partial_{\rho}(\rho B_{\varphi})
-  B_{\varphi} \partial_{\varphi}(B_{\rho}) -  B_{\rho} \partial_{\rho}(\rho B_{\rho}) - B_{\rho} \partial_{\varphi}(B_{\varphi}) \right]  \\
& =  \left[ B_{\varphi} \partial_{\rho}(\rho B_{\varphi}) -  B_{\rho} \partial_{\rho}(\rho B_{\rho}) - \partial_{\varphi}(B_{\rho} B_{\varphi}) \right].
\end{split}
\end{equation}

\noindent Therefore, one can obtain 

\begin{equation}\label{VMF}
\begin{split}
& \left[\partial_{\rho}(\rho B_{\varphi}) - \partial_{\varphi}(B_{\rho})\right] B_{\varphi}  =  \left[ B_{\varphi}^{2} - B_{\rho}^{2} \right]+ \\
& \left[(\frac{\rho}{2}) \left[ \partial_{\rho}(B_{\varphi}^{2}) 
-  \partial_{\rho}(B_{\rho}^{2}) \right]- \partial_{\varphi}(B_{\rho} B_{\varphi})\right].
\end{split}
\end{equation}

\noindent What we need is a mean radial dependence, so we apply the average $ \left\langle \; \right\rangle\equiv \frac{1}{4 \pi h} \int_{0}^{2 \pi}  \int_{-h}^{h} dz d\varphi $ to Eq. \;(\ref{VMF}) \citep{Elstner2014}:

\begin{equation}\label{VMFAV}
\begin{split}
&\left\langle v_{mag}^{2} \right\rangle  = \frac{1}{4\pi \varrho}\left[ \left( \left\langle B_{\varphi}^{2}\right\rangle  - \left\langle B_{\rho}^{2}\right\rangle\right)  + \dfrac{\rho}{2} \left( \partial_{\rho} \left\langle B_{\varphi}^{2}\right\rangle -  \partial_{\rho} \left\langle  B_{\rho}^{2} \right\rangle \right) \right] \\
& - \frac{1}{4\pi \varrho} \left[ \frac{\rho}{4 \pi h} \int_{0}^{2 \pi} \left[ B_{\rho}(h) B_{z}(h) - B_{\rho}(-h) B_{z}(-h) \right] d\varphi \right].\\
& + \frac{1}{4\pi \varrho}\left[ \frac{\rho}{2} \partial_{\rho} \left\langle B_{z}^{2}\right\rangle  \right] 
\end{split}
\end{equation}

\noindent Note that by applying the average to Eq. \;(\ref{VMF}), the last term has vanished, since  $\int_{0}^{2 \pi}  \partial_{\varphi}(B_{\rho} B_{\varphi}) d \varphi = B_{\rho} B_{\varphi}\vert_{0}^{2 \pi} = 0$.\\
%$ \left\langle \partial_{\varphi}(B_{\rho} B_{\varphi}) \right\rangle =  \int_{0}^{2 \pi}  \partial_{\varphi}(B_{\rho} B_{\varphi}) d \varphi = B_{\rho} B_{\varphi}\vert_{0}^{2 \pi} = 0$.
\noindent Usually, the regular component of the magnetic field used to be considered purely horizontal and mainly parallel to the galactic plane ($B_{z} = 0$) \citep[e.g.,][]{ Ruiz-Granados2010, Ruiz-Granados2012, Battaner2007, Battaner2000}, since the large-scale magnetic fields of spiral galaxies have a dominant azimuthal component $B_{\varphi}$, which are high enough that they should be taken into account when modeling the rotation curve of spiral galaxies.\\
\noindent Observations of the large-scale poloidal X-shaped magnetic fields in many spiral galaxies, if viewed externally and edge-on, are indications of the large-scale vertical magnetic fields, $B_{z}$ \citep{Henriksen2016, Nixon2018, Woodfinden2019}.
Magnetic field transport from the disk of the nearby spiral galaxy $NGC\;6946$ into the halo \citep{Heald2012}. 
Therefore to estimate the possible contribution of such magnetic field on the dynamical motion of the gaseous distribution of the galactic disk and hence on the rotation curves,  the whole three-dimensional structure of the magnetic field should be taken into account.\\
\noindent For a differentially rotating disk-like galaxy, mean-field dynamo theory predicts a quadrupolar magnetic field configuration, where the azimuthal magnetic field direction is the same above and below the plane, but the direction of the vertical component of the magnetic field reverses with respect to the plane \citep{Haverkorn2012}. Therefore, we can consider a symmetric magnetic field with respect to the galactic plane and set the vertical magnetic field at the disk surface:
\begin{equation}\label{eqBz}
B_{z}(h) = - B_{z}(-h)
\end{equation}
\noindent which is also clear from Eq. \ref{down}.
Hence, we can rewrite Eq. \ref{VMFAV} as:
\begin{equation}\label{VMFAV2}
\begin{split}
&\left\langle v_{mag}^{2} \right\rangle  = \frac{1}{4\pi \varrho}\left[ \left( \left\langle B_{\varphi}^{2}\right\rangle  - \left\langle B_{\rho}^{2}\right\rangle\right)  + \dfrac{\rho}{2} \left( \partial_{\rho} \left\langle B_{\varphi}^{2}\right\rangle -  \partial_{\rho} \left\langle  B_{\rho}^{2} \right\rangle \right)\right] \\
&+ \frac{1}{4\pi \varrho}\left[ \frac{\rho}{2} \partial_{\rho} \left\langle B_{z}^{2}\right\rangle - \frac{ \rho}{4 \pi h} \int_{0}^{2 \pi} B_{z}(h) \left[ B_{\rho}(h) + B_{\rho}(-h) \right] d\varphi \right].
\end{split}
\end{equation}

Since, the magnetic field is divergence free, $\nabla \cdot B = 0$, therefore,
\begin{equation}\label{div}
\frac{1}{\rho} \partial_{\rho}(\rho B_{\rho}) + \frac{1}{\rho} \partial_{\varphi}( B_{\varphi}) + \partial_{z}( B_{z}) = 0
\end{equation}
By applying the average $ \left\langle \; \right\rangle\equiv \frac{1}{4 \pi h} \int_{0}^{2 \pi}  \int_{-h}^{h} dz d\varphi $
to Eq. \ref{div},

\begin{equation}
\frac{1}{\rho} \partial_{\rho}(\rho \langle B_{\rho}\rangle) + \frac{1}{4 \pi h}  \int_{0}^{2 \pi}  \int_{-h}^{h}\partial_{z}( B_{z}) dz d\varphi = 0
\end{equation}

\begin{equation}
\frac{1}{\rho} \partial_{\rho}(\rho \langle B_{\rho}\rangle) + \frac{1}{2 h}  B_{z} \vert_{-h}^{h} = 0
\end{equation}

\noindent Therefore
\begin{equation}
 B_{z}(h) = -  B_{z}(- h) = -  \frac{h} {\rho}  \partial_{\rho}(\rho \langle B_{\rho}\rangle) 
\end{equation}
Note that by applying the average to Eq.\;(\ref{div}), the second term has vanished, since $ \int_{0}^{2 \pi}  \partial_{\varphi} B_{\varphi} d \varphi =  B_{\varphi}\vert_{0}^{2 \pi} = 0$.

\noindent From Eq. \ref{up}
\begin{equation}\label{eqBrho}
B_{\rho}(z = h) = B_{\rho}(z = -h)
\end{equation}

\noindent Hence, we can rewrite Eq. \ref{VMFAV} as:
\begin{equation}\label{VMFAV2Final}
\begin{split}
\left\langle v_{mag}^{2} \right\rangle & =   \frac{1}{4\pi \varrho}\left[ \left( \left\langle B_{\varphi}^{2}\right\rangle  - \left\langle B_{\rho}^{2}\right\rangle\right)  + \dfrac{\rho}{2} \left( \partial_{\rho} \left\langle B_{\varphi}^{2}\right\rangle -  \partial_{\rho} \left\langle  B_{\rho}^{2} \right\rangle \right)\right] \\
 & + \frac{1}{4\pi \varrho}\left[  \frac{\rho}{2} \left(  \partial_{\rho} \left\langle B_{z}^{2}\right\rangle - \frac{1}{\pi h} \int_{0}^{2 \pi} \left[ B_{\rho}(h) B_{z}(h) \right] d\varphi \right) \right].
\end{split}
\end{equation}

\subsection{Two main magnetic spiral arms in galaxy $NGC\;6946$ and a superposition of $m = 0$ and $m = 2$ modes }
%{\bf The observational values of linear polarization for NGC 6946 are taken from \citep{Beck2007}.}
\noindent One classic example of disk-dominated magnetic fields can be seen in the spiral galaxy $NGC\;6946$. Magnetic fields in the disks of spiral galaxies have a spiral morphology that is strongly correlated with the orientation and pitch angle of traditional tracers of gravitational spiral arms, such as massive stars and dust lanes \citep{Braun2010}.

\noindent The spiral galaxy $NGC\;6946$ hosts two surprisingly symmetric magnetic spiral arms, with greater symmetry than the optical arms, highly aligned magnetic fields, which are located between the gas/optical arms \citep{Beck2007, beckandHoernes1996, Becketal1996}. 
The global magnetic field of a spiral galaxy is observed to be either axisymmetric (azimuthal mode $m = 0$) or bisymmetric (mode $m = 1$) in its azimuthal component \citep{vallee1991}. 
Polarized radio observations of the spiral galaxy $NGC\;6946$ showed that its regular magnetic field configuration with a dominant two-armed pattern cannot be explained with simple ASS ($m=0$ mode), BSS ($m=1$ mode), or QSS ($m=2$ mode) models \citep{Beck2007, beckandHoernes1996, Becketal1996, Rohde-Beck-1999, Arshakian2007}. 
\cite{Woodfinden2019} noted that the m = 0 mode is required to produce the X-shaped magnetic fields.\\
A good model for describing two main magnetic spiral arms in the gas-rich galaxy $NGC\;6946$ should be consistent with the magnetic field configuration, strength, and direction of the field.\\
In the spiral galaxy $NGC\;6946$, the strengths of the ordered, mostly regular magnetic field in the two main magnetic spiral arms (which are strongest at around $5 \; kpc$ radius) is typical $8-10 \; \mu G$ \citep{Beck2007}.
The two main magnetic spiral arms in this galaxy are more than $12 \; kpc$ long and only about $0.5-1 \; kpc$ wide and do not fill all of the interarm regions, unlike the polarized radio emission in M81 \citep{Beck2007, beckandHoernes1996, Becketal1996}. The $m = 1$ dynamo mode generates a bisymmetric two-armed spiral structure but of a much larger width than the two main inner magnetic arms which are discovered in galaxy $NGC\;6946$ \citep{beckandHoernes1996}. 
The Bisymmetric model ($m = 1$ dynamo mode) leads to two dominant magnetic spiral arms with opposite field directions of the radial field component with respect to the galaxy's center, while the spiral field in both of the main inner magnetic spiral arms in $NGC\;6946$ point inwards (directed towards the galaxy's center).
The regular magnetic fields of spiral shape can be represented as a superposition of modes with different azimuthal symmetries.
The two bright inner magnetic spiral arms in $NGC\;6946$, with the field directed towards the galaxy's center in both, can be interpreted as a superposition of two even azimuthal dynamo modes, $m = 0$ and $m = 2$, since the superposition of even azimuthal modes $m = 0$ and $m = 2$ with the same peak amplitudes, may lead to two dominant magnetic spiral arms with the same direction of the radial field component with respect to the galaxy's center \citep{Beck2007, beckandHoernes1996, Becketal1996, Rohde-Beck-1999, vallee1991, Arshakian2007}, while the superposition of the azimuthal modes $m = 0$ and $m = 1$ with the same peak amplitudes, lead to only one dominant magnetic spiral arm, as has been shown in Fig.\;(\ref{m=0,1,2}a), and the superposition of the azimuthal modes $m = 1$ and $m = 2$ with the same peak amplitudes, lead to one main magnetic spiral arm and two secondary magnetic spiral arms, as has been shown in Fig.\;(\ref{m=0,1,2}b). In Fig.\;(\ref{m=0,1,2}b), the direction of the radial field component of the main magnetic arm is the opposite of the direction of the radial field component of two secondary magnetic spiral arms, with respect to the galaxy's center. The magnetic field lines of the main magnetic arm have a clockwise direction, while those of the two secondary magnetic spiral arms have a counterclockwise direction (see Fig.\;\ref{m=0,1,2}b).

\section{Modeling the Rotation Curve}\label{Modeling the Rotation Curve}

\noindent A galaxy is a system of stars, interstellar gas, dark matter halo, and magnetic fields, that affect the rotation of the gas and stars through the magnetic force. The rotation curve of disk-like galaxies is the main observable kinematic data, allowing the investigation of the dynamical properties of their stars and interstellar gas \citep{Sofue2001}.
In this section, considering the three-dimensional structure for the regular magnetic field, for two dark matter halo models, ISO and NFW profile, we model the rotation curve by means of the contribution of different components, an exponential stellar disk, an exponential gas disk, a dark matter halo model, and the regular magnetic field, to describe the observed rotation curve of $NGC\;6946$.
The gravitational rotation velocity is split into contributions from stellar, gaseous disks, dark matter, and magnetic field.
The total circular velocity is obtained by quadratic summation of the different contributions: 
\begin{equation}
v_{c}^2=\sum_{i}^{n}v_{c}(i)^2.
\end{equation}
\noindent So, by considering the magnetic field effects and taking its contribution into account to the dynamics, the total circular velocity is obtained as: 
\begin{equation}\label{eqtotal}
v_{c}^2=v_{stellar\hspace{0.1cm} disk}^{2} + v_{gas\hspace{0.1cm} disk}^{2} + v_{dark\hspace{0.1cm} matter}^2 + v_{magnetic}^2 \;,
\end{equation}
\noindent where $v_{stellar\hspace{0.1cm} disk}$, $v_{gas\hspace{0.1cm} disk}$ and $v_{dark\hspace{0.1cm} matter}$ are the stellar disk, gas disk, and dark matter halo contribution to the circular velocity, respectively. The contribution to the circular velocity of the regular magnetic field has been shown by $v_{magnetic}$.

\subsection{Baryonic Contribution and the Exponential Disk Model} 

\noindent The surface density profiles of galactic disks are usually exponential (\citealt{binbook, Dutton2009}). 
\begin{equation}
\mu_{r} = \mu_{0} \exp\left( - \dfrac{r}{R_{exp}}\right) 
\end{equation}
\noindent where $\mu_{0}$ is the central surface density and $R_{exp}$ is the exponential disk scale length.
The stellar disk and gas disk contribution to the circular velocity is derived using the potential of such an exponential mass distribution.
The potential that such an exponential disk would generate is given by \cite{binbook}:
\begin{equation}\label{eqpoten}
\begin{split}
 \Phi_{Disk}(r,0) =& - 4G\mu_{0}\int_{0}^{r}d a \frac{a K_{1}(a/R_{exp})}{\sqrt{r^2-a^2}}=\\
 &-\pi G \mu_{0}r\left[I_{0}(y)K_{1}(y)-I_{1}(y)K_{0}(y)\right]
 \end{split}
\end{equation}
\noindent with
\begin{equation}
y \equiv \frac{r}{2 R_{exp}}.
\end{equation}
\noindent The exponential disk contribution to the circular velocity is obtained with  differentiating Eq. \eqref{eqpoten} with respect to $r$:
\begin{equation}
v_{disk}^2=4\pi G \mu_{0} R_{exp}y^{2}\left[I_{0}(y)K_{0}(y)-I_{1}(y)K_{1}(y)\right]
\end{equation}

where $K_{n}$ and $I_{n}$ are modified Bessel functions.

\subsection{DARK MATTER CONTRIBUTION}

\noindent Here we consider and use two well-known models for the dark matter halo distribution in our analysis, which are described by ISO and  NFW profiles.\\

%\begin{itemize}
%\item {\bf ISO profile}
%\end{itemize}
{\bf ISO profile}\\

\noindent The Isothermal profile is the representative model of a core-like halo with solid-body behavior \citep{Begeman1991} and describes a description of the distribution of dark matter in (gas-rich, late-type) dwarfs and disk galaxies which are dominated by dark matter.
The ISO profile of dark matter is as follows:
\begin{equation}
\rho=\frac{\rho_{0}}{1+\left(\frac{r}{R_{c}}\right)^{2}}
\end{equation}
with the free parameters, $ \rho_{0} $ and $ R_{c}$, being the central density and the core radius of the halo.
The circular velocity is as follows:
\begin{equation}
v_{dark\hspace{0.1cm} matter}^{2}=4\pi G \rho_{0}R_{c}^{2}\left[1-\frac{R_{c}}{r}arctan\left(\frac{r}{R_{c}}\right)\right].
\end{equation}\\

%\begin{itemize}
%\item {\bf NFW profile}
%\end{itemize}
{\bf NFW profile}\\

\noindent The NFW profile, which is known as the universal profile, is the commonly accepted DM halo profile in the standard CDM cosmology  \citep{NFW1996}. This profile was obtained from the collisionless $N$ body simulations of the clustering of dark matter particles by \cite{NFW1997}.
The density profile is given by:
\begin{equation}
\rho_{NFW}=\frac{\rho_{c}}{\left(\frac{r}{R_{s}}\right)\left(1+\frac{r}{R_{s}}\right)^{2}}.
\end{equation}
The NFW profile has two free parameters, the representative density $\rho_{c}$ and the scale radius $R_{s}$ ( the characteristic radius of the halo) \citep{NFW1997}. 
The corresponding gravitational potential for this density profile, NFW, is:
\begin{equation}
\Phi_{NFW}(r) = - \frac{4 \pi G \rho_{c} R_{s}^{3}}{r} \ln \left(1 +  \frac{r}{R_{s}}\right).
\end{equation}
The contribution to the circular velocity due to the NFW profile of dark matter is as follows:
\begin{equation}
v_{dark\hspace{0.1cm} matter}^{2}=\frac{4\pi G R_{s}^{3}\rho_{c}}{r}\left[\ln\left(1+\frac{r}{R_{s}}\right)-\frac{r}{r+R_{s}}\right]
\end{equation}
where $ \rho_{c} $ is related to the density of the universe at the time of halo collapse and $ R_{s} $ is the characteristic radius of the halo.

\subsection{Best-fit solutions for the rotation curve}\label{Best Fit}

\noindent Our analysis is based on a reduced $\chi^{2}$, an $\chi$-squared minimization method (least-squares fitting) as the goodness-of-fit statistic. For galaxy $NGC\;6946$, we perform fitting to the observational data points, with two mass density profiles, ISO and NFW for the dark matter halo, with and without considering the contribution of the regular magnetic field to the circular velocity.
For every mass model, the contribution of the dark matter halo is determined, with two free parameters, $\rho_{0}$ and $R_{c}$ for ISO profile, $\rho_{0}$ and $a$ for NFW profile. For the gas  and stellar disk, the different parameters are considered as fixed
values (see Table \ref{tab1}) in our analysis. The contribution to the rotation curve of three-dimensional regular magnetic field has been modeled as a combination of the planar magnetic field and an out-of-plane quadrupolar poloidal field component. The planar magnetic field topology is a superposition of two azimuthal dynamo modes $m = 0$ (Axisymmetric Model, Eq.\;\ref{ASS}) and $m = 2$ (Quadri-symmetric Spiral Model, Eq.\;\ref{2-+}).

\noindent The contribution of the regular magnetic field to the rotation curve has been also fitted through a free parameter, $\rho_{1}$, which has been already defined in Eq.\;(\ref{r1}).
So, without taking into account the contribution of the regular magnetic field to the dynamics, the model rotational velocity, which should be fitted to the observed rotational velocity, is a function of two free parameters (i.e., dark matter halo parameters), which represent the contribution of the dark matter halo term to the rotation curve.
Taking into account the contribution of the regular magnetic field to the dynamics, the model rotational velocity is a function of three free parameters, two halo parameters, and $\rho_{1}$, where the last parameter represents the contribution of the perturbation of the regular magnetic field to the rotation curve.\\
The parameters for best fit to the observed rotation curve are obtained by minimizing the reduced $\chi^{2}$ function:
\begin{equation}\label{chy}
\chi_{r}^{2}=\frac{1}{(N - M)}\sum_{i=1}^{N}\left[\frac{v_{c, the}(r_{i}) - v_{c, obs}(r_{i})}{\sigma_{i}}\right]^2.
\end{equation}
\noindent In this equation, $ v_{c, obs} (r_{i})$ is the observed circular velocity related with $i$-th data point at radius $r_{i}$ and $v_{c, the}$ is the model circular velocity (defined by Eq. \ref{eqtotal}) for this data point and $\sigma_{i}$ is the observational error bar related with each data point. We have used the data points from the observational high-resolution rotation curve of $NGC\;6946$ from The HI Nearby Galaxy Survey (THINGS), presented by \cite{deBlok2008}, which are the highest quality HI rotation curves available to date for a large sample of nearby galaxies. 
In Eq. \ref{chy}, N is the total number of data points from the observed rotation curve, and $M $ is the number of free parameters
($M = 2$ without the contribution of the regular magnetic field, and $M = 3$ with the contribution of the regular magnetic field).\\
Table\;(\ref{tab1}) contains fixed parameters for the gaseous disk, the stellar disk, and the regular magnetic field for $NGC\;6946$. Column (1) gives the exponential stellar disk scale length, $R_{exp, \star}$, and column (2) gives the exponential gaseous disk scale length, $R_{exp, gas}$. Column (3) gives the central surface density for a stellar disk, $ \mu_{0, \star} $, and column (4) gives the central surface density for the gaseous disk, $ \mu_{0, gas}$. The gaseous disk mainly contains atomic gas, $HI$, molecular gas, $H_{2}$, and helium. Column (5) gives the mean pitch angle of the northern main magnetic arm, $p_{north}$.  Finally, column (6) gives the mean pitch angle of the southern main inner magnetic arm structure, $p_{south}$ and column (7) gives the vertical exponential scale height of the radio halo with a thin radio disk for a typical spiral galaxy with X-shaped magnetic field structure.

\begin{figure*}[hbt!]
\centering
	\includegraphics[width=1.1\columnwidth]{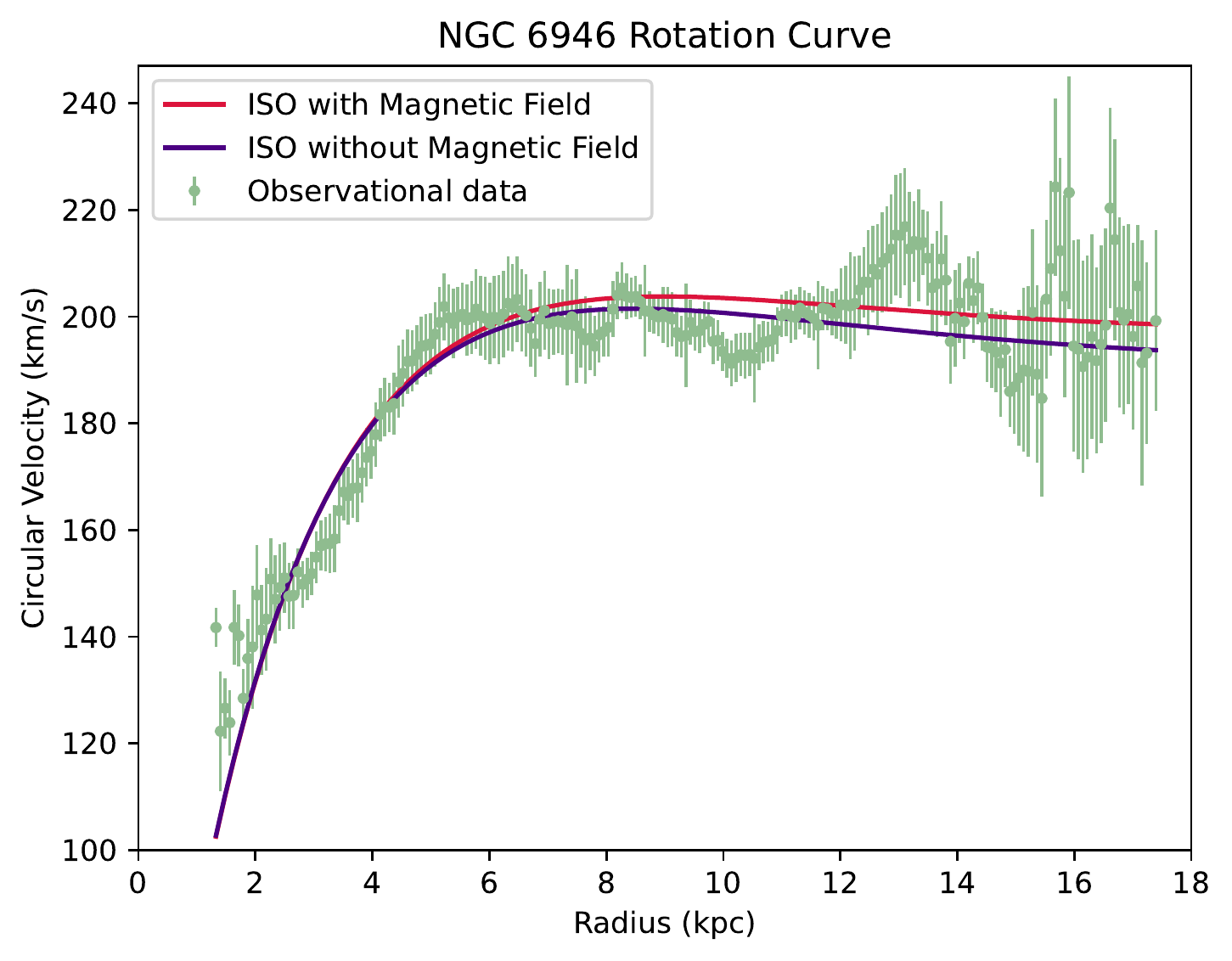}
    \caption{Best-fit solutions for the rotation curve of $NGC\;6946$, with and without including the contribution of the magnetic component with Isothermal-Sphere Halos.}
    \label{figISO}
\end{figure*}

\begin{figure*}[hbt!]
 \centering
	\includegraphics[width=1.1\columnwidth]{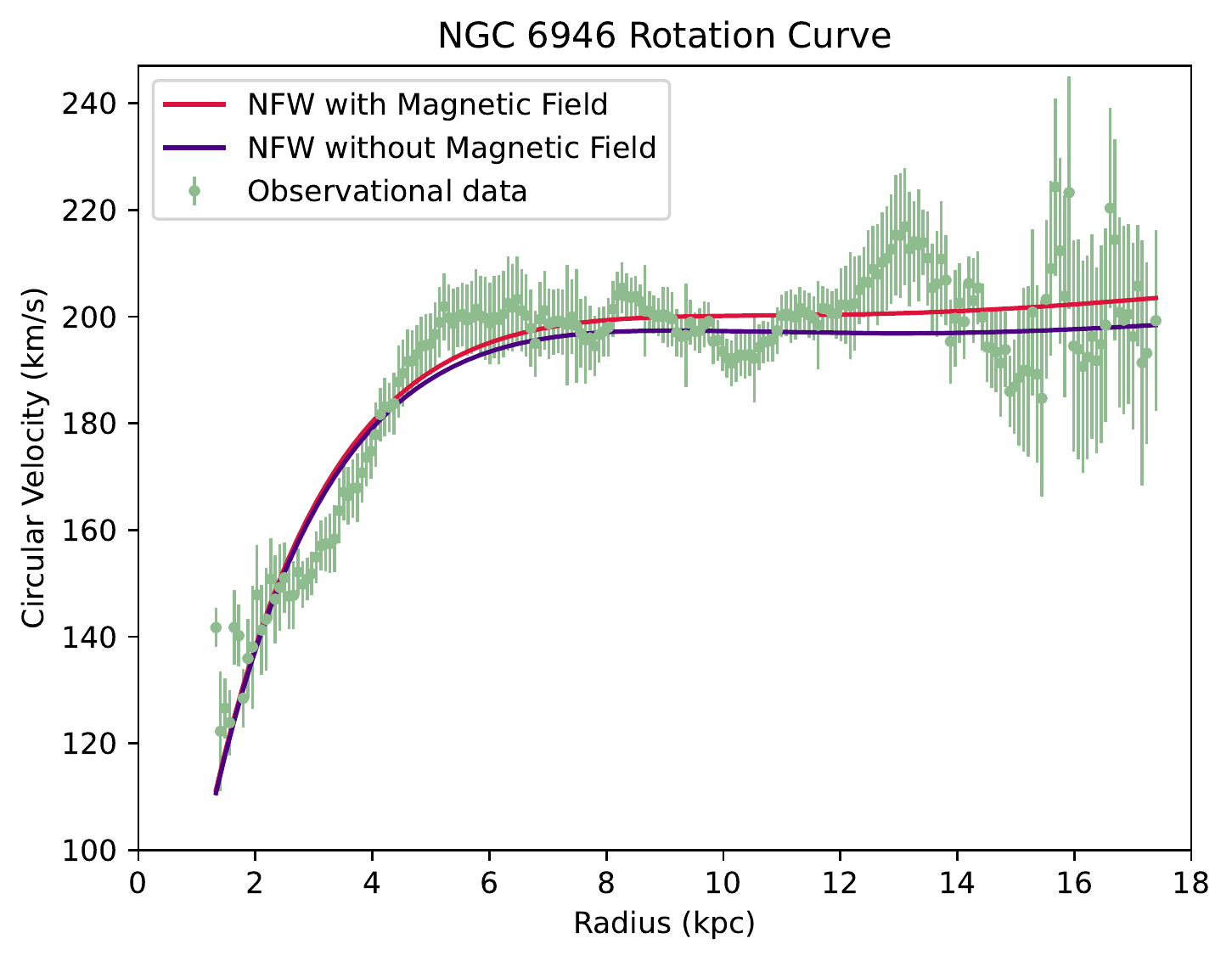}
    \caption{Best-fit solutions for the rotation curve of $NGC\;6946$, with and without including the contribution of the magnetic component with NFW profile.}
    \label{figNFW}
\end{figure*}

\begin{table*}[hbt!]
%\centering
\caption{Fixed Parameters for the Gaseous Disk, the Stellar Disk, and the Regular Magnetic Field for $NGC\;6946$}
\label{tab1}
\begin{threeparttable}
\begin{tabular}{ccccccccccccc}
\hline
(1)&(2)&(3)&(4)&(5)&(6)&(7)\\
\hline
$R_{exp, \star}$ & $R_{exp, gas}$ & $ \mu_{0, \star} $ & $ \mu_{0, gas}$ & $p_{north}$ & $p_{south}$ & $h_{exp}$ &&&\\
(kpc) & (kpc) & ($M_{Sun}/pc^{2})$ &  ($M_{Sun}/pc^{2}) $ & (degree)  & (degree) & (kpc) &&&\\
 \hline
 2.57 \tnote{a} & 4.9 \tnote{a} & 900 \tnote{a} &  0.16 \tnote{a} & $-25^{\degree} \pm 1^{\degree}$ \tnote{b} &  $-31^{\degree}  \pm 2^{\degree}$ \tnote{b} & 0.3 \tnote{c, d ,e, f} &&& \\
  \hline
 \end{tabular}
   \begin{tablenotes}
  \item\label{3} References: (a) \cite{Casasola2017}, (b) \cite{Beck2007}, (c) \cite{Dumke1998}, (d) \cite{Krause2009}, (e) \cite{Heesen2008}, (f) \cite{Soida2011}.
  \end{tablenotes}
\end{threeparttable}
\end{table*}
By programming in python, we find the best-fit solutions for the Isothermal-Sphere halo profile and NFW profile. Our main results are summarized in Fig.\;(\ref{figISO}) for the Isothermal-Sphere halo profile and in Fig.\;(\ref{figNFW}) for the NFW profile. In each of these figures, the red line and the blue line show the corresponding best fit to the observed rotation curve, with and without including the contribution of the magnetic component in the dynamic for galaxy $NGC\;6946$, respectively.

Table (\ref{tab2}) summarizes our results for the best-fit solutions for the Isothermal-Sphere halo profile and NFW profile, which shows the constrained parameters for two different cases i.e., without including the contribution of the magnetic component and by taking into account the magnetic term in the dynamic. 
\begin{table*}[hbt!]
%\centering
\caption{Best-fits Solutions to the Rotation Curve for the ISO and NFW Profiles, without and with the Contribution of Magnetic Fields}
\label{tab2}
\begin{threeparttable}
\begin{tabular}{ccccccccc}
\hline
Model &&&& Derived Parameters &&&& $\chi_{r}^{2}$ for Best-fit \\
\hline
\hline
ISO &&&& $ \rho_{0} =  5.77 \times 10 ^{-2}$  ($M_{Sun}/pc^{2}$) &&&& 1.8 \\
&&&& $ R_{c}$ = 3.45 (kpc) &&&&  \\
\hline
ISO + Magnetic Field &&&& $ \rho_{0} =  5.58 \times 10 ^{-2} $ ($M_{Sun}/pc^{2}$) &&& & 1.6 \\
&&&& $ R_{c} $ = 3.66 (kpc)  &&&&\\
&&&& $ \rho_{1} $ = 9 (kpc)  &&&&\\
\hline
\hline
NFW &&&&  $\rho_{c}$ = $ 2.35 \times 10^{-3} $($M_{Sun}/pc^{2}$) &&&& 1.4  \\
&&&& $R_{s}$ = 41.8 (kpc) &&&& \\
\hline
NFW + Magnetic Field &&&& $\rho_{c}$ = $ 2.25 \times 10^{-3}  $ ($M_{Sun}/pc^{2}$) &&&& 1.1 \\
&&&& $R_{s}$ = 45.2 (kpc) &&&& \\
&&&& $ \rho_{1} $ = 9 (kpc) & &&&\\
 \hline
 \end{tabular}
\end{threeparttable}
\end{table*}

\begin{figure*}[hbt!]
\centering
	\includegraphics[width=.90\columnwidth]{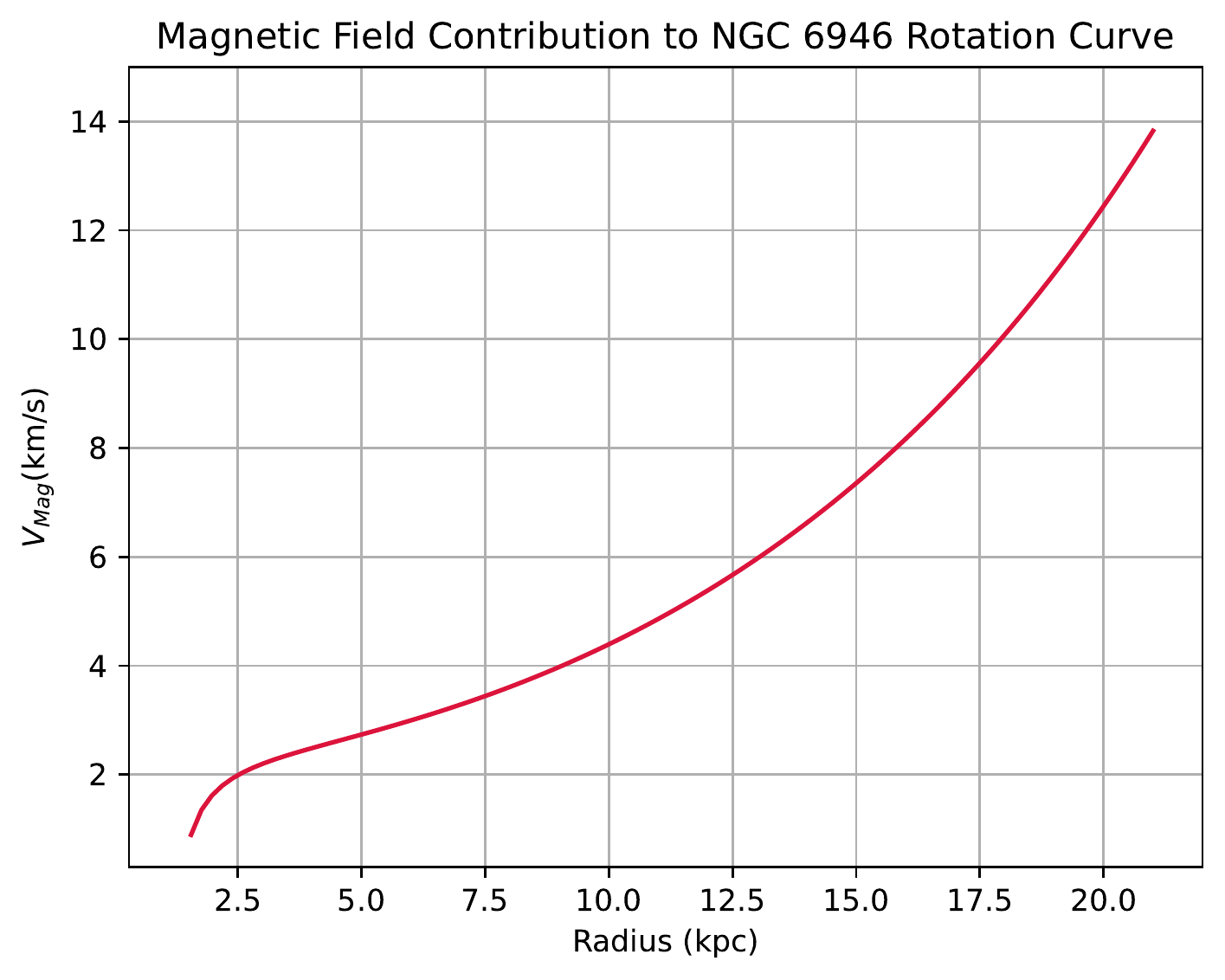}
    \caption{Contribution of the modeled regular magnetic field caused by the two main inner arms to $NGC\;6946$ rotation curve.}
    \label{MagneticFieldContribution}
\end{figure*}

\noindent Our results suggest that regular magnetic field strength at $\rho_{1} \sim 9 \; kpc $ is about half its value around the galactic center. This value agrees with those observations derived from polarized radio synchrotron emission and Faraday rotation measurements (RM)\citep{Beck2007}.
For the modeled regular magnetic field for the spiral galaxy $NGC\;6946$, which is a superposition of two azimuthal dynamo modes $m = 0$ (axisymmetric model) and $m = 2$ (quadri-symmetric spiral model), with the best-fit solution for the characteristic scale radius of the regular magnetic field strength, $\rho_{1}$ ($\rho_{1} \sim \; 9 kpc$), 
we plot the contribution of the regular magnetic field to the $NGC\;6946$ circular velocity in terms of radius in Fig.\;(\ref{MagneticFieldContribution}), which shows an ascending curve with a typical amplitude of $ 6 - 14 \;  km s^{-1}$ in the outer gaseous disk of the galaxy $NGC\;6946$.

We also plot the contribution ratios of the regular magnetic field caused by the two main magnetic inner arms in the rotation curve, to the observed circular velocity, i.e., $v_{B}/v_{Ob}$ and to the dark matter contribution, i.e., $v_{B}/v_{DM}$ in terms of radius in Fig. \ref{ratio}. The contribution ratio of the regular magnetic field to the observed circular velocity increases with the galactocentric radius and is about five percent in the outer regions of the galaxy $NGC\;6946$.

\begin{figure*}[hbt!]\label{ratio}
    \subfigure[]{ \includegraphics[width=0.64\columnwidth]{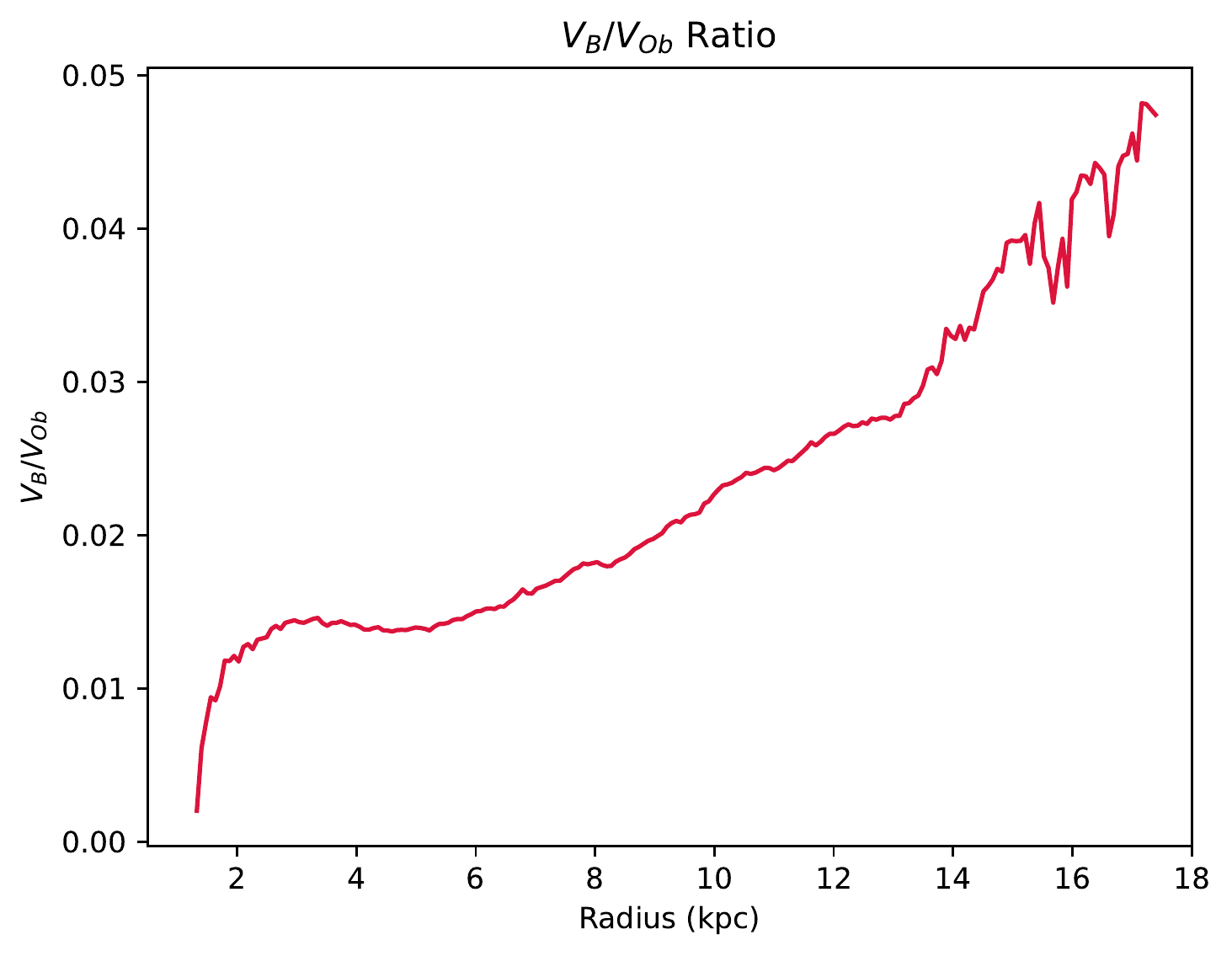}}\quad
       \subfigure[]{ \includegraphics[width=0.64\columnwidth]{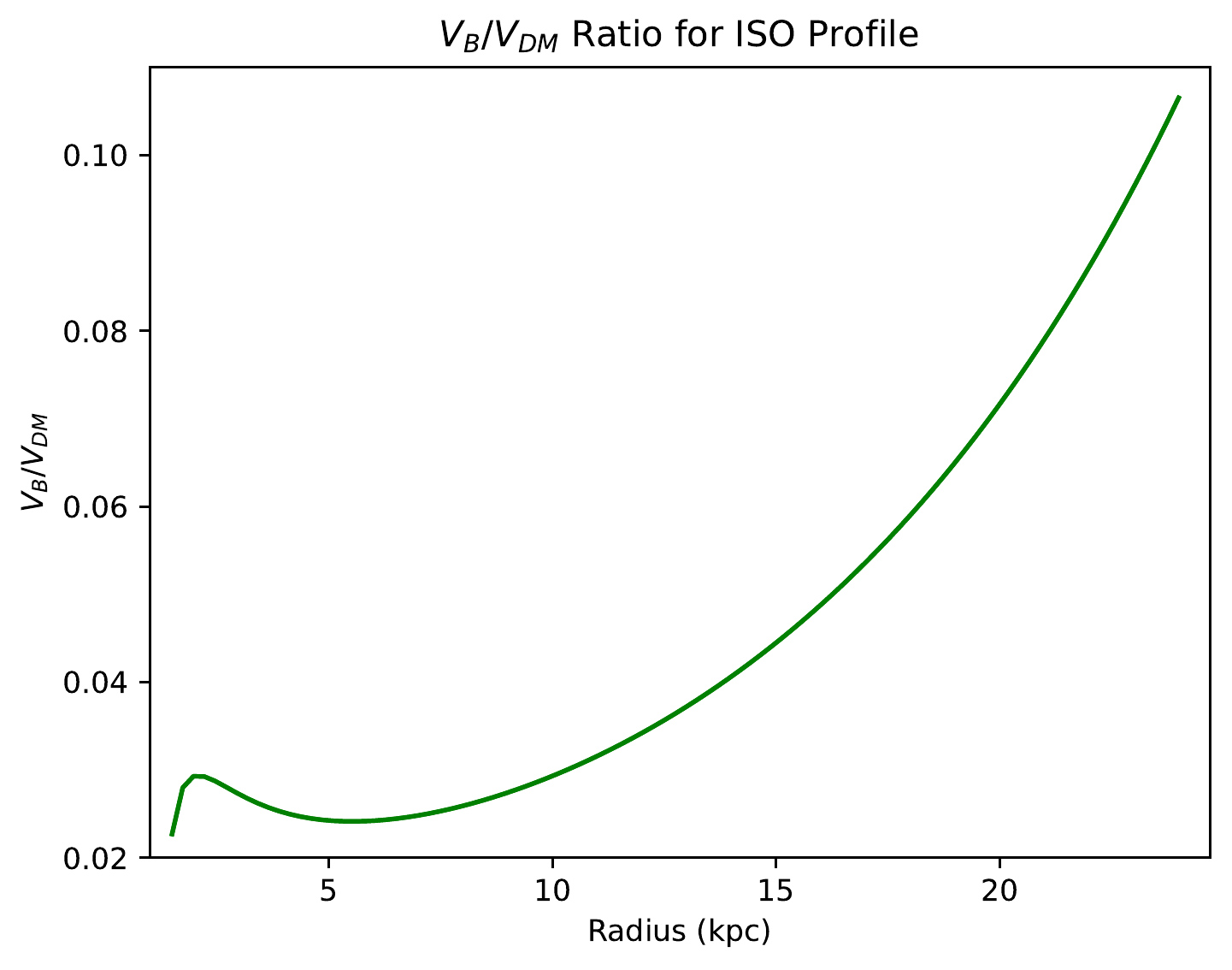}}\quad
       \subfigure[]{ \includegraphics[width=0.64\columnwidth]{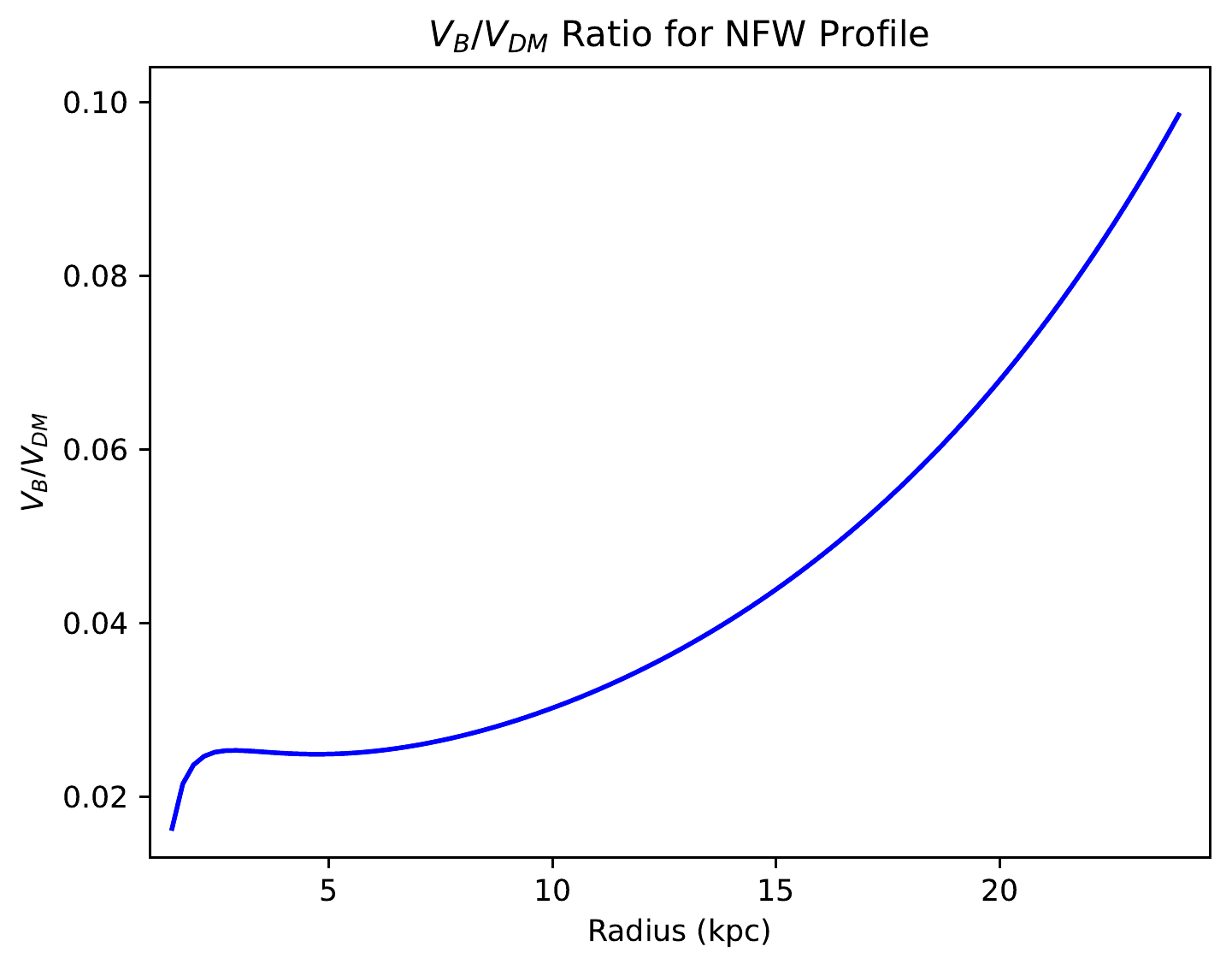}}\\
    \caption{(a) Contribution ratio of the regular magnetic field caused by the two main inner arms to the observed circular velocity, $v_{B}/v_{Ob}$. (b and c) The contribution ratio of the regular magnetic field to dark matter in the rotation curve, $v_{B}/v_{DM}$ in terms of radius for (b) ISO profile and (c) NFW model.}
\label{ratio}
\end{figure*}

\noindent Then by using the derived parameter for the regular magnetic field contribution, i.e., the characteristic scale radius, we generate the map of the modeled regular magnetic field strength for the spiral galaxy $NGC\;6946$ in Fig.\;(\ref{m=0,2}), which is a superposition of two azimuthal dynamo modes $m = 0$ (axisymmetric model) and $m = 2$ (quadri-symmetric spiral model) with the same amplitudes and the average pitch angle $p_{avg} = - 28^{\degree}$, which is the average of the mean pitch angles of the northern and southern main inner magnetic arms. This map clearly shows two main inner contours with the spiral pattern and the highest magnetic field strength, with the field directed towards the galaxy's center in both, almost compatible with two inner main bright magnetic arms detected in the polarized synchrotron intensity maps \citep{beckandHoernes1996, Beck2007, BeckandWielebinski2013, Beck2001, Becketal1996}.

\begin{figure*}[hbt!]
\centering
	\includegraphics[width=1.3\columnwidth]{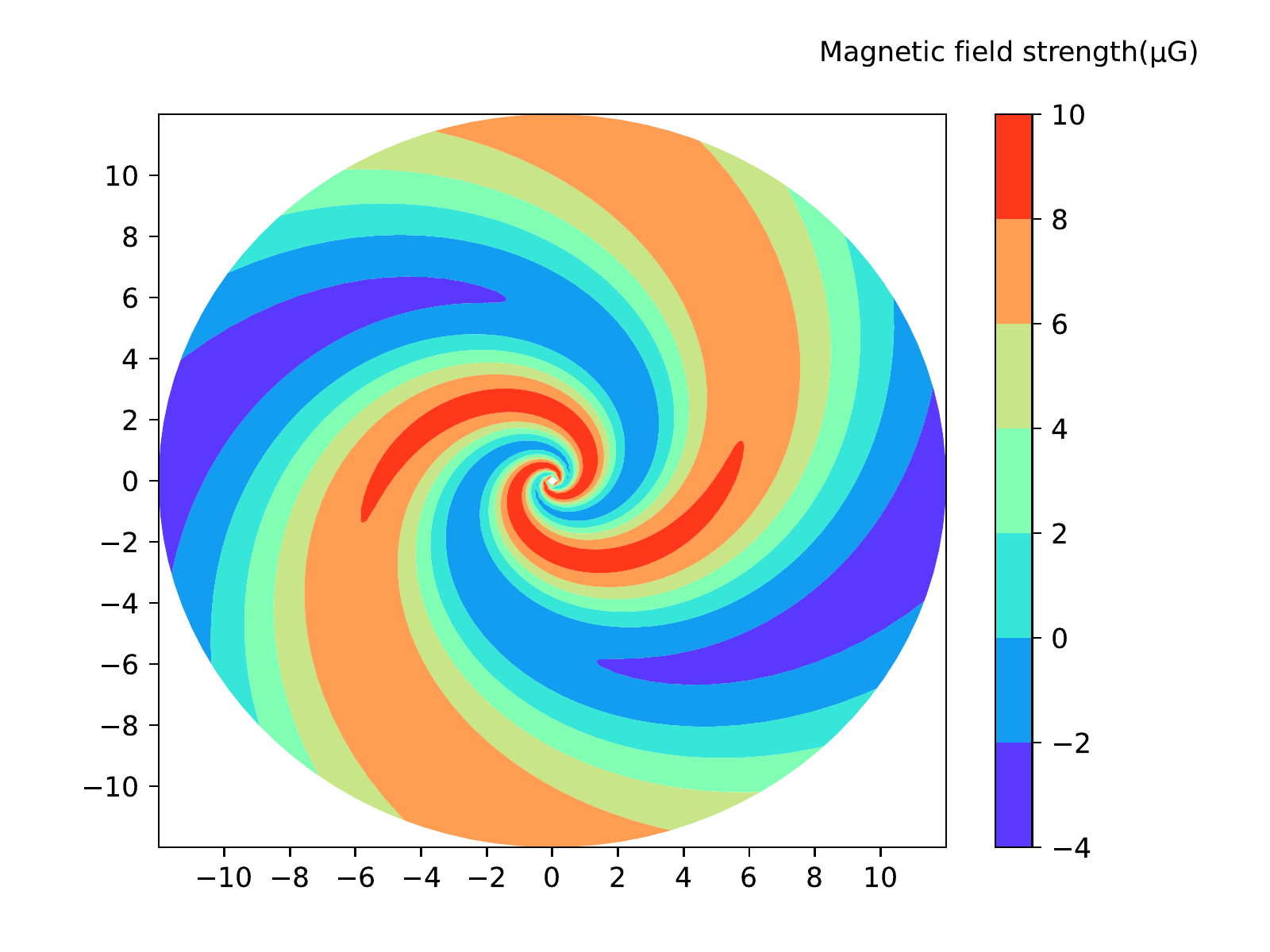}
    \caption{The map of the planar component of the modeled regular magnetic field strength for the spiral galaxy $NGC\;6946$ in the galactic disk plane (in the $X-Y$ plane), extracted from a superposition of two azimuthal dynamo modes $m = 0$ (axisymmetric model) and $m = 2$ (quadri-symmetric spiral model) with the same amplitudes, which leads to two dominant magnetic spiral arms with the same direction of the radial field component with respect to the galaxy's center (same field direction). The pitch angle is considered to be the average of the mean pitch angles of the northern and southern main inner magnetic arms, $p_{avg} = - 28^{\degree}$.}
    \label{m=0,2}
\end{figure*}

\section{Discussion}\label{Dis}

\noindent  Our result of best fitting for the characteristic scale radius, $\rho_{1}$, suggests that regular magnetic field strength at $\rho_{1} \sim 9 \; kpc$  is about half its value around the galactic center. This value agrees with those observations derived from polarized radio synchrotron emission and Faraday rotation measurements (RM)\citep{Beck2007}.
\noindent By using the derived parameter for the regular magnetic field contribution, i.e., the characteristic scale radius, $\rho_{1}$, we also constructed the map of the planar component of the modeled regular magnetic field strength for the spiral galaxy $NGC\;6946$ in Fig.\;\ref{m=0,2}, which is a superposition of two azimuthal dynamo modes $m = 0$ (axisymmetric model) and $m = 2$ (quadri-symmetric spiral model). The structure of the extracted map clearly shows two main inner contours with the highest magnetic field strength,  $8-10 \; \mu G$, and the field directed towards the galaxy$^{,}$s center in both, almost compatible with two inner main bright magnetic arms, which have been revealed in the polarized synchrotron intensity maps \citep{beckandHoernes1996, Beck2007, BeckandWielebinski2013, Beck2001, Becketal1996}. \\
\noindent In an attempt to explain the rising of the rotation curve in the outer disks of the Milky Way and M 31, \cite{Ruiz-Granados2010, Ruiz-Granados2012} considered a regular magnetic field, purely azimuthal (toroidal) and proportional to 1/R, described by an axisymmetric model ($m = 0$ azimuthal dynamo mode). They concluded that the contribution of the regular magnetic field is positive and able to rise up the rotation curve in the outer parts of the disk. The radial decrease of the large-scale magnetic field strength is slower than that of gravity since gravity decreases as ${1}/{R^{2}}$, but the magnetic field decreases as ${1}/{R}$. Therefore the gas rotation may be influenced by the magnetic fields, especially in the outer regions \citep{Battaner2000, Battaner2007}.\\
\noindent \cite{Elstner2014} assumed an exponential profile for the regular magnetic field to estimate the contribution of the radial component of the Lorentz force and magnetic field to the gas rotation in several spiral galaxies.
They found that the magnetic field may have a positive contribution within a radius $\rho_{B_{reg}}$ of about $ 20 kpc$ from the center and a possible negative contribution to the rotation curve in the outermost parts of galaxies. $\rho_{B_{reg}}$ is the scale length of the regular magnetic field.\\
\noindent In this paper, we modeled the regular magnetic field structure of the spiral galaxy $NGC\;6946$ in three-dimension by considering spiral magnetic field patterns as a superposition of two even azimuthal dynamo modes, $m = 0$ and $m = 2$, nearly parallel in the galactic plane, and modifying it with a poloidal X-shaped above and below the plane. Then by fitting the modeled rotation curve to the observed data, we found that the contribution of the magnetic field to the galactic dynamics can lead to increasing rotational velocity at large radii (see Fig. \ref{MagneticFieldContribution}).
The azimuthal component of the regular magnetic field caused by the two main inner arms contributes to an inward-directed Lorentz force, but the radial component contributes to an outward-directed Lorentz force (see Eq. \ref{VMFAV}). Even though the magnetic force from the azimuthal field will be reduced by the contribution of the radial component of the regular magnetic field and the negative contribution of the last term in Eq. \ref{VMFAV2Final}, i.e., $ - 2 \int_{0}^{2 \pi} \left[ B_{\rho}(h) B_{z}(h) \right] d\varphi $ for an X-shaped magnetic field structure, our modeled regular magnetic field has a positive contribution in the rotation velocity of the $HI$ gas.\\
\noindent We found a positive contribution of the magnetic field in the rotational velocity, which agrees with \cite{Ruiz-Granados2010, Ruiz-Granados2012} since they found such a positive contribution. However, their model is much simpler than ours. They considered the regular magnetic field to be purely azimuthal in the plane of galaxies, but we modeled the regular magnetic field in three dimensions.
Our result for the contribution of the magnetic field in the rotation curve of spiral galaxy $NGC\;6946$ disagrees with the results of \cite{Elstner2014}.  We found a positive contribution for the regular magnetic field, especially at the outermost region, while \cite{Elstner2014} found a negative contribution for it. The reason is the difference between our model and their model. They considered an exponential radial profile for the magnetic field.\\ 
The reason why the sign of contribution of the magnetic field in the rotational velocity is different for different profiles of the regular magnetic field is the dependency of $v^{2}_{mag}$  on the magnetic field configuration. The contribution of the large-scale (ordered) magnetic fields to the rotation curve, $v^{2}_{mag}$, depends on the magnetic field configuration \citep{Sanchez-Salcedo2013}. 
\cite{Sanchez-Salcedo2013} showed that the sign of $v^{2}_{mag}$ depends on the model of magnetic field structure; for example for an exponential profile, $v^{2}_{mag}$ is negative at $\rho > \rho_{B}$ and is positive at $\rho < \rho_{B}$, which $\rho_{B}$ is the scale length of the regular magnetic field. But for the axisymmetric model ($B_{\varphi} \propto {1}/{R}$), $v^{2}_{mag}$ is positive at any radius.\\
It has been suggested that if the contribution of magnetic fields leads to an increase in the rotational velocity, values of the mass-to-light ratio are more realistic in the disks of the galaxies $NGC 891$ and $ NGC 253$ \citep{Jalocha2012}.
\cite{Sanchez-Salcedo2004} found that the magnetic field contribution cannot speed up $HI$ discs by more than $20 \;  km s^{-1}$ at the outermost regions of HI detection.
We found that the contribution of the modeled regular magnetic fields caused by the two main inner arms to the rotation velocity of the $HI$ gas shows an ascending curve with a typical amplitude of about $6 - 14 \;  km s^{-1}$ in the outer gaseous disk of the galaxy $NGC\;6946$ (see Fig.\;\ref{MagneticFieldContribution}).
The contribution ratio of the regular magnetic field to the observed circular velocity and to dark matter increases with the galactocentric radius. Its ratio to the observed rotational velocity is about five percent and to dark matter is about 10 percent in the outer regions of the galaxy $NGC\;6946$ (see Fig. \ref{ratio}). Therefore the magnetic field contribution to the galactic dynamics becomes more important at large radii in the outer region of the galaxy.

\section{SUMMARY}\label{Conclusion}
\noindent To study the dynamical effect of the galactic disk magnetic fields on the rotation curve of the spiral galaxy $NGC\;6946$, we modeled a three-dimensional magnetic field and its contribution to the rotation curve.
We performed fitting via an $\chi$-squared minimization method to the observational data points, with two mass density profiles, ISO and the NFW for the dark matter halo, for two different cases i.e., with and without considering the contribution of the regular magnetic field to the circular velocity.
For both mass models, by taking into account the contribution of the regular magnetic field to this galaxy dynamics, the fit is improved in the outer part of this galaxy, and magnetic effects on the gaseous disk decrease the value of the reduced $\chi^{2}$ statistic. 
The contribution of the regular magnetic fields in the rotation velocity of the $HI$ gas increases with the galactocentric radius and has a typical amplitude of about $ 6 - 14 \;  km s^{-1}$ in the outer regions of the galaxy $NGC\;6946$ (see Fig.\;\ref{MagneticFieldContribution}). The contribution ratio of the regular magnetic field to dark matter in the observed circular velocity is about 10 percent at large radii in the outer regions. Even though large-scale magnetic fields cannot be an alternative to dark matter, it still has their own contribution to the large-scale dynamics of spiral galaxies, especially in the outer parts of galaxies.

%\bigskip
%\bigskip
 %\newpage
 
\section*{Acknowledgments}
We would like to thank the anonymous referee for the very useful remarks and constructive comments that have substantially improved the manuscript.
We are very grateful to Prof. Erwin de Blok for providing us with the data on the rotation curves of the THINGS galaxies that are presented in their paper \citep{deBlok2008}.

\bigskip

%\appendix

%\section{Explicit Calculation}
%\pagebreak 
%\newpage
 
\bibliography{M.Khademi}{}
\bibliographystyle{aasjournal}



\end{document}